\documentclass[aps,epsfig,manuscript]{revtex4-1}
\usepackage{amsmath,amssymb,graphicx}
\usepackage{epsfig}
\usepackage{graphicx}
\usepackage{enumerate}
\usepackage{natbib}

\newcommand{\beq}{\begin{equation}}
\newcommand{\eeq}{\end{equation}}

\newcommand{\bea}{\begin{eqnarray}}
\newcommand{\eea}{\end{eqnarray}}

\begin{document}



\title
{Path-integral simulations with fermionic and bosonic reservoirs:\\
Transport and dissipation
in molecular electronic junctions}
\author{Lena Simine}
\affiliation{Chemical Physics Theory Group, Department of Chemistry, University of Toronto,
80 Saint George St. Toronto, Ontario, Canada M5S 3H6}
\author{Dvira Segal}
\affiliation{Chemical Physics Theory Group, Department of Chemistry,
University of Toronto, 80 Saint George St. Toronto, Ontario, Canada
M5S 3H6}

\date{\today}
\begin{abstract}
We expand iterative numerically-exact influence functional
path-integral tools and present a method capable of following the
nonequilibrium time evolution of subsystems coupled to multiple
bosonic and fermionic reservoirs simultaneously.
Using this method, we study the real-time dynamics of charge
transfer and vibrational mode excitation in an electron conducting
molecular junction. We focus on nonequilibrium
vibrational effects, particularly, the development of vibrational
instability in a current-rectifying junction.
Our simulations are performed by assuming large molecular
vibrational anharmonicity (or low temperature). This allows us to
truncate the molecular vibrational mode to include only a two-state
system.
Exact numerical results are compared to perturbative
Master equation calculations demonstrating an excellent agreement
in the weak electron-phonon coupling regime. Significant deviations
take place only at strong coupling.
Our simulations allow us to quantify the contribution of different
transport mechanisms, coherent dynamics and inelastic transport, in
the overall charge current. This is done by studying two
model variants: The first admits inelastic electron transmission
only, while
the second one allows for both coherent and incoherent pathways.
\end{abstract}

\maketitle


\section{Introduction}
\label{Sec1}


Following the quantum dynamics of an open-dissipative many-body system
with multiple bosonic and fermionic reservoirs in a nonequilibrium
state, beyond the linear response regime, is a significant theoretical and
computational challenge.
In the realm of molecular
conducting junctions, we should describe the
out-of-equilibrium dynamics of the molecular unit while handling
both electrons and molecular vibrations, accounting for
many-body effects such as electron-electron, phonon-phonon and
electron-phonon interactions.
Given this complexity, studies in this field are mostly focused on
steady-state properties, using e.g., scattering theory
\cite{Scatt1,Scatt2,Scatt3}, while ignoring vibrational
nonequilibrium effects. Perturbative treatments (in either the
molecule-leads coupling parameter or the electron-phonon interaction
energy) are commonly used, including the nonequilibrium Green's
function technique \cite{NitzanVib,Millis,Galperin1,Galperin2,Jonas}
and Master equation approaches
\cite{Millis,HeatSegal,Wege,Thoss1,Thoss2,Peskin}. For following the
{\it real-time dynamics} of such systems, involved methods have been
recently developed, e.g., semiclassical approaches
\cite{Swenson1,Swenson2}.

\begin{figure}
\hspace{0.5mm} \vspace{-4mm}
\includegraphics[width=.42\textwidth, angle=270]{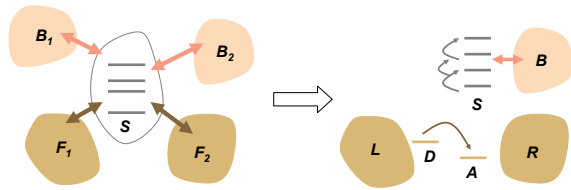}
\vspace{-22mm}
\caption{Left Panel: Generic setup considered in this work,
including a subsystem ($S$) coupled to multiple
fermionic ($F$) and bosonic ($B$) reservoirs. Right panel: Molecular
electronic realization with two metals, $L$ and $R$, connected
by two electronic levels, $D$ and $A$. Electronic transitions in this
junction are coupled to excitation/de-excitation processes of a
particular, anharmonic, vibrational mode that plays the role of the
``subsystem". This mode may dissipate its excess energy to a
secondary phonon bath $B$.} \label{Fig:fig0}
\end{figure}


In this work, we extend numerically-exact
path-integral methods, and  follow the dynamics of a
subsystem coupled to multiple out-of-equilibrium bosonic and
fermionic reservoirs. The technique is then applied on a molecular
junction realization, with the motivation to address basic problems
in the field of molecular electronics. Particularly, in
this work we consider the dynamics and steady-state properties of a conducting
molecular junction acting as a charge rectifier. A scheme of the
generic setup and a particular molecular junction realization are
depicted in Fig. \ref{Fig:fig0}.

The time evolution scheme developed in this paper treats both
bosonic and fermionic reservoirs. This is achieved by combining two
related iterative path-integral methods: (i) The quasi-adiabatic
path-integral approach (QUAPI) of Makri et al.
\cite{QUAPI1,*QUAPI2}, applicable for the study of subsystem-boson
models, and (i) the recently developed influence-functional
path-integral (INFPI) technique \cite{IF1,*IF2}, able to produce the
dynamics of subsystems in contact with multiple fermi baths. The
latter method (INFPI) essentially generalizes QUAPI. It relies on
the observation that in out-of-equilibrium (and/or finite
temperature) situations bath correlations have a finite range,
allowing for their truncation beyond a memory time dictated by the
voltage-bias and the temperature. Taking advantage of this fact, an
iterative-deterministic time-evolution scheme can be developed,
where convergence with respect to the memory length can in principle
be reached.

The principles of the INFPI approach have been detailed in
\cite{IF1,*IF2}, where it has been adopted for investigating
dissipation effects in the nonequilibrium spin-fermion model and
charge occupation dynamics in correlated quantum dots. Recently, it
was further utilized for examining the effect of a magnetic flux on
the intrinsic coherence dynamics in a double quantum dot system
\cite{AB}, and for studying relaxation and equilibration dynamics in
finite metal grains \cite{Kunal1,*Kunal2}.


Numerically-exact methodologies are typically limited to simple
models; analytic results are further restricted to specific
parameters. The Anderson-Holstein (AH) model has been studied
extensively in this context. In this model the electronic structure
of the molecule is represented by a single spinless electronic
level, with electron occupation on the dot coupled  to
the displacement  of a single oscillator mode, representing an internal
vibration. This vibration may connect with  a secondary phonon bath,
representing a larger phononic environment (internal modes,
solvent). The AH model has been simulated exactly with the
secondary phonon bath, using a a real-time path-integral Monte Carlo
approach \cite{Rabani}, and by extending the multilayer
multiconfiguration time-dependent Hartree method to include
fermionic degrees of freedom \cite{ThossExact}. More recently, the
model has been simulated by adopting the iterative-summation of
path-integral approach \cite{egger1,egger2,eggerP}.

In this paper, we examine a variant of the AH model, the Donor
(D)-Acceptor (A) electronic rectifier model \cite{Aviram}. This
model incorporates {\it nonlocal} electron-vibration interactions:
electronic transitions between the two molecular states, A and D,
are coupled to a particular internal molecular vibrational mode.
Within this simple system, we are concerned with the development of
vibrational instability: Significant molecular heating can take
place once the D level is lifted above the A level, as the excess
electronic energy is used to excite the vibrational mode. This
process may ultimately lead to junction instability and breakdown
\cite{Lu}. We have recently studied a variant of this model
(excluding direct D-A tunneling element), using a Master equation
method, by working in the weak electron-phonon coupling limit.
\cite{ET}.
An important observation in that work has been that since the
development of this type of instability is directly linked to the
breakdown of the detailed balance relation above a certain bias
(resulting in an enhanced vibrational excitation rate constant, over
relaxation), it suffices to describe the vibrational mode as a
truncated two-level system. In this picture, population inversion in
the two-state system evinces on the development of vibrational
instability.

Our objectives here are threefold: (i) To present a numerically-exact
iterative scheme for following the dynamics of a quantum
system driven to a nonequilibrium steady-state due to its coupling
to multiple bosonic and fermionic reservoirs. (ii) To demonstrate
the applicability of the method in the field of molecular
electronics. Particularly, to explore the development of vibrational
instability in conducting molecules.
(iii) To evaluate the performance and accuracy of
standard-perturbative Master equation treatments, by comparing their
predictions to exact results. Since Master equation techniques are
extensively used for explaining charge transfer phenomenology,
scrutinizing their validity and accuracy is an important task.

The plan of the paper is as follows. In Sec. \ref{Sec2} we introduce
the path-integral formalism. We describe the iterative time
evolution scheme in Sec. \ref{Sec3}, by exemplifying it to the case
of a spin subsystem. Sec. \ref{Sec4} describes a molecular
electronics application, and we follow both electrons and
vibrational dynamics in a dissipative molecular rectifier. Sec.
\ref{Sec5} concludes. For simplicity, we use the conventions
$\hbar\equiv 1$, electron charge $e\equiv 1$, and Boltzmann constant
$k_B=1$.


\section{Path-integral formulation}
\label{Sec2}

We consider a multi-level subsystem, with the Hamiltonian $H_S$,
coupled to multiple bosonic ($B$) and fermionic ($F$) reservoirs that
are prepared in an out-of-equilibrium initial state.
The total Hamiltonian $H$ is written as
\bea
H=H_S+H_{B}+H_F+ V_{SB}+V_{SF}.
\label{eq:HP}
\eea
In the energy representation of the isolated subsystem, its Hamiltonian can be written as
\bea
H_S=\sum_{s}\epsilon_s|s\rangle \langle s| + \sum_{s\neq s'}v_{s,s'}|s\rangle \langle s'|.
\label{eq:HSP}
\eea
The Hamiltonian $H_F$ may comprise of multiple fermionic baths, and
similarly, $H_B$ may contain more than a single bosonic reservoir.
The terms $V_{SF}$ and $V_{SB}$ include the coupling of the
subsystem to the fermionic and bosonic environments, respectively.
Coupling terms which directly link
the subsystem to both bosonic and fermionic degrees of freedom are not included. However,
$V_{SB}$ and $V_{SF}$ may contain non-additive contributions with
their own set of reservoirs. For example, $V_{SF}$ may admit
subsystem assisted tunneling terms, between separate fermionic baths
(metals), see Fig. \ref{Fig:fig0}.

We are interested in the time evolution of the reduced density
matrix $\rho_S(t)$.
This quantity is obtained by tracing the total density matrix $\rho$
over the bosonic and fermionic reservoirs' degrees of freedom
\bea
\rho_S(t) = {\rm Tr}_B{\rm Tr}_F \left[ e^{-iHt}\rho(0)e^{iHt}\right].
\label{eq:rhoSP}
\eea
We also study the dynamics of certain expectation values,
for example, charge current and  energy current. The time
evolution of an operator $A$ can be calculated using the relation
\bea \langle  A(t) \rangle &=& {\rm Tr} [\rho(0)  A(t)]
\nonumber\\
&=&
\lim_{\lambda \rightarrow 0} \frac{\partial}{\partial \lambda} {\rm
Tr}\big[\rho(0) e^{iHt}e^{\lambda A}e^{-iHt} \big].
\label{eq:AtP}
\eea
Here $\lambda$ is a real number, taken to vanish at the end of the
calculation. When unspecified, the trace is performed over
the subsystem states and all the environmental degrees of freedom. In what follows, we
detail the path-integral approach for the calculation of the reduced
density matrix. Section \ref{Sec3e} presents expressions
useful for time-evolving expectation values of operators.

As in standard path-integral approaches, we decompose the time
evolution operator into a product of $N$ exponentials,
$e^{iHt}=\left(e^{iH\delta t}\right)^N$ where $t=N\delta t$, and
define the discrete time evolution operator $\mathcal{G}\equiv
e^{iH\delta t}$. Using the Trotter decomposition,
we approximate $\mathcal G$ by
\bea
\mathcal{G}\sim  \mathcal{G_F} \mathcal{G_B} \mathcal{G_S}\mathcal{ G_B}\mathcal{ G_F},
\label{eq:trotter1P}
\eea
where we define
\bea
\mathcal{G_F}&\equiv&e^{i(H_F+V_{SF})\delta t/2}, \,\,\,\,
\mathcal{G_B}\equiv e^{i(H_B+V_{SB})\delta t/2}
\nonumber\\
\mathcal{G_S}&\equiv&e^{iH_S\delta t}
\label{eq:trotter2P}
\eea
Note that the breakup of the subsystem-bath term,
$e^{i(H_F+V_{SF}+H_B+H_{SB})\delta t/2} \sim
\mathcal{G_B}\mathcal{G_F}$, is exact if the commutator
$[V_{SB},V_{SF}]$ vanishes. This fact allows for an exact separation between
the bosonic and fermionic influence functionals, as we explain
below. This commutator nullifies if the fermionic and bosonic baths
couple to commuting subsystem degrees of freedom, for example,
$V_{SB}\propto |s\rangle \langle s|$ and $V_{SF}\propto |s'\rangle
\langle s'|$.

As an initial condition, we assume that at time $t=0$ the
subsystem and the baths are decoupled, $\rho(0)=\rho_S(0) \otimes
\rho_B\otimes \rho_F$, and the baths are prepared in a
nonequilibrium (biased) state. For example, we may include in
$H_{F}$ two Fermi seas that are prepared each in a grand canonical
state with different chemical potentials and temperatures.
%
The overall time evolution can be represented by a path-integral
over the subsystem states,
\begin{widetext}
\bea
&&\langle s_N^+| \rho_S(t)| s_N^- \rangle
\nonumber\\
&&=\sum_{s^{\pm}_0}\sum_{s_1^{\pm}}... \sum_{s_{N-1}^{\pm}}
{\rm Tr}_B{\rm Tr}_F \Big[
\langle s_N^+|\mathcal G^{\dagger}|s^+_{N-1}\rangle
\langle s_{N-1}^+|\mathcal G^{\dagger}|s^+_{N-2}\rangle...
\langle s_{0}^+|\rho(0)|s^-_{0}\rangle ...
\langle s_{N-2}^-|\mathcal G|s^-_{N-1}\rangle
\langle s_{N-1}^-|\mathcal G|s_N^-\rangle
 \Big].
\nonumber\\
\label{eq:pathP}
\eea
Here $s_k^{\pm}$ represents the discrete path on the forward ($+$)
and backward ($-$) contour. The calculation of each discrete
term is done by introducing four additional summations,
e.g.,
\bea \langle s_{k}^-|{\mathcal G}|s_{k+1}^-\rangle=
\sum_{f^{-}_k}\sum_{g^{-}_k}\sum_{m^{-}_k}\sum_{n^{-}_k} \langle
s_{k}^-|\mathcal{G_F} |f_k^-\rangle \langle f_k^-| \mathcal{G_B}
|m_k^-\rangle \langle m_k^-| \mathcal{G_S} |n_k^-\rangle \langle
n_k^-|\mathcal{ G_B} |g_k^-\rangle \langle g_k^-| \mathcal{G_F}
|s_{k+1}^-\rangle.
\nonumber\\
 \label{eq:GP} \eea
\end{widetext}
We substitute Eq. (\ref{eq:GP}) into Eq. (\ref{eq:pathP}), further
utilizing the factorized subsystem-reservoirs initial condition as
mentioned above, and find that the function under the sum can be
written as a product of separate terms,
\bea \langle s_N^+|\rho_S(t)|s_N^-\rangle = \sum_{\bf s^{\pm}}
\sum_{\bf f^{\pm}} \sum_{\bf g^{\pm}} \sum_{\bf m^{\pm}} \sum_{\bf
n^{\pm}}
I_S({\bf m^{\pm}},{\bf n^{\pm}},s^{\pm}_0)
I_F({\bf s'^{\pm}},{\bf f^{\pm}},{\bf g^{\pm}})
I_B({\bf f^{\pm}},{\bf m^{\pm}},{\bf n^{\pm}}, {\bf g^{\pm}}).
\nonumber\\
\label{eq:IBFSP}
\eea
Here $I_S$ follows the subsystem ($H_S$) free
evolution. The  term $I_F$ is referred to as a fermionic ``influence
functional" (IF), and it contains the effect of the fermionic
degrees of freedom on the subsystem dynamics. Similarly, $I_B$, the
bosonic IF, describes how the bosonic degrees of freedom affect the
subsystem. Bold letters correspond to a path, for example, ${\bf
m^{\pm}}=\{m_0^{\pm},m_1^{\pm},...,m_{N-1}^{\pm}\}$. We also define
the path ${\bf s^{\pm}}=\{s_0^{\pm},s_1^{\pm},...,s_{N-1}^{\pm}\}$,
and the associate path which covers $N+1$ points, ${\bf
s'^{\pm}}=\{s_0^{\pm},s_1^{\pm},...,s_{N-1}^{\pm}, s_{N}^{\pm}\}$.
Given the product structure of Eq. (\ref{eq:IBFSP}), the subsystem,
bosonic and the fermionic terms can be independently evaluated,
while coordinating their path. Explicitly, the elements in Eq.
(\ref{eq:IBFSP}) are given by
\bea
I_S&=&\langle s_0^+| \rho_S(0)|s_0^{-}\rangle \Pi_{k=0,...,N-1}
\langle m_k^-|{\mathcal G_S}|n_k^-\rangle
\langle n_k^+|{\mathcal G_S^{\dagger}}|m_k^+\rangle
\nonumber\\
I_F&=&
{\rm Tr}_F \Big[\langle s_N^+|{\mathcal G_F^{\dagger}}|g_{N-1}^+\rangle
\langle f_{N-1}^+|{\mathcal G_F^{\dagger}}|s_{N-1}^+\rangle...
\nonumber\\
&\times&
\langle s_1^+|{\mathcal G_F^{\dagger}}|g_{0}^+\rangle
\langle f_0^+|{\mathcal G_F^{\dagger}}|s_{0}^+\rangle
\rho_F
\langle s_0^-|{\mathcal G_F}|f_{0}^-\rangle
\langle g_0^-|{\mathcal G_F}|s_{1}^-\rangle...
\nonumber\\
&\times&
\langle s_{N-1}^-|{\mathcal G_F}|f_{N-1}^-\rangle
\langle g_{N-1}^-|{\mathcal G_F}|s_{N}^-\rangle \Big]
\nonumber\\
I_B&=&
{\rm Tr}_B \Big[\langle g_{N-1}^+|{\mathcal G_B^{\dagger}}|n_{N-1}^+\rangle
\langle m_{N-1}^+|{\mathcal G_B^{\dagger}}|f_{N-1}^+\rangle...
\nonumber\\
&\times&
\langle g_0^+|{\mathcal G_B^{\dagger}}|n_{0}^+\rangle
\langle m_0^+|{\mathcal G_B^{\dagger}}|f_{0}^+\rangle
\rho_B
\langle f_0^-|{\mathcal G_B}|m_{0}^-\rangle
\langle n_0^-|{\mathcal G_B}|g_{0}^-\rangle...
\nonumber\\
&\times& \langle f_{N-1}^-|{\mathcal G_B}|m_{N-1}^-\rangle \langle
n_{N-1}^-|{\mathcal G_B}|g_{N-1}^-\rangle \Big]. \eea
The dynamics in Eq. (\ref{eq:IBFSP}) can be retrieved by following an
iterative scheme, by using the principles of the INFPI approach
\cite{IF1,*IF2}. In the next section we illustrate this evolution with
a spin subsystem.

\section{Iterative time evolution scheme}
\label{Sec3}

We consider here the spin-boson-fermion model. It includes a
two-state subsystem that is coupled through its polarization to
bosonic and fermionic reservoirs. With this relatively simple model,
we exemplify the iterative propagation technique, see Secs.
\ref{Sec3a}-\ref{Sec3e}. Relevant expressions for a multi-level
subsystem and general interaction form are included in Sec.
\ref{Sec3f}.

\subsection{spin-boson-fermion model}
\label{Sec3a}
The spin-fermion model, with a qubit, spin, coupled to a fermionic bath
 is kindred to the eminent spin-boson model, describing a qubit interacting with bosonic
environment. It is also related to the Kondo model \cite{Kondo}, only
lacking direct coupling of the reservoir degrees of freedom to spin-flip processes.
It provides a minimal setting for the study of dissipation and decoherence effects
in the presence of nonequilibrium reservoirs \cite{MitraSpin1, MitraSpin2, SMarcus,SF}.
Here we put together the spin-boson and the spin-fermion models, and present it in the general
form,
\bea
H_S&=&\Delta \sigma_x + B \sigma_z,\,\,\,\,\,
\nonumber\\
H_F&=&\sum_{j}\epsilon_j c_j^{\dagger} c_j
+ \sum_{j\neq j'}v_{j,j'}^F c_{j}^{\dagger} c_{j'}
\nonumber\\
V_{SF}&=&\sigma_z\sum_{j,j'}\xi^F_{j,j'}c_{j}^{\dagger}c_{j'}.
\nonumber\\
H_B&=&\sum_p \omega_p b_p^{\dagger}b_p + \sum_{p,p'} v_{p,p'}^B b_p^{\dagger}b_{p'},
\nonumber\\
V_{SB}&=&\sigma_z\sum_p\xi_{p}^B \left(b_p^{\dagger}+b_p \right)  +
\sigma_z\sum_{p,p'}\zeta_{p,p'}^B b_p^{\dagger}b_{p'}.
\label{eq:HS}
\eea
The subsystem includes only two
states, with an energy gap $2B$ and a
tunneling splitting  $2\Delta$. This minimal subsystem is coupled
here through its polarization to a set of boson and fermion degrees
of freedom, where
$\sigma_{z}$ and $\sigma_{x}$
denote the $z$ and $x$ Pauli matrices for a two-state subsystem, respectively.
$b_p$ stands for a bosonic operator, to destroy a mode of frequency
$\omega_p$, similarly, $c_j$ is a fermionic operator, to annihilate an electron
of energy $\epsilon_j$ (we assume later a linear dispersion relation).
 In this model, spin polarization couples to
harmonic displacements, to scattering events between electronic
states in the metals (fermi reservoirs), and to scattering evens
between different modes in the harmonic bath. Since the commutator
between the interaction terms vanish, $[V_{SF},V_{SB}]=0$, the
separation between the bosonic and fermionic IFs is exact. Moreover,
since the fermionic and bosonic operators couple both to $\sigma_z$,
we immediately note that $f_k^{\pm}=s_k^{\pm}$,
$m_k^{\pm}=f_k^{\pm}$, $n_k^{\pm}=g_k^{\pm}$ and
$g_k^{\pm}=s_{k+1}^{\pm}$. Eq. (\ref{eq:IBFSP}) then  simplifies to
\bea
\langle s_N^+|\rho_S(t)|s_N^-\rangle =
\sum_{\bf s^{\pm}}
I_S({\bf s'^{\pm}})
I_F({\bf s'^{\pm}})
I_B({\bf s'^{\pm}}),
\label{eq:IBFSS}
\eea
where we recall the definitions of the paths ${\bf
s^{\pm}}=\{s_0^{\pm},s_1^{\pm},...,s_{N-1}^{\pm}\}$ and ${\bf
s'^{\pm}}=\{s_0^{\pm},s_1^{\pm},...,s_{N-1}^{\pm}, s_{N}^{\pm}\}$.
The subsystem evolution and the IFs are now given by
\begin{widetext}
\bea &&I_S({\bf s'^{\pm}})=
 \langle s_0^+ | \rho_S(0)|s_0^- \rangle
K(s_N^{\pm},s_{N-1}^{\pm})... K(s_2^{\pm},s_1^{\pm})K(s_1^{\pm},s_0^{\pm})
\nonumber\\
&&I_B({\bf s'^{\pm}})={\rm Tr}_B\Big[
e^{-iW_B(s_N^+) \delta t/2} e^{-iW_B(s_{N-1}^+) \delta t}...
 e^{-iW_B(s_{0}^+) \delta t/2} \rho_B e^{iW_B(s_0^-) \delta
t/2}....
e^{iW_B(s_{N-1}^-) \delta t} e^{iW_B(s_{N}^-) \delta t/2} \Big].
\nonumber\\
&&
I_F({\bf s'^{\pm}})={\rm Tr}_F\Big[
e^{-iW_F(s_N^+) \delta t/2} e^{-iW_F(s_{N-1}^+) \delta t}...
 e^{-iW_F(s_{0}^+) \delta t/2} \rho_F e^{iW_F(s_0^-) \delta
t/2}....
e^{iW_F(s_{N-1}^-) \delta t} e^{iW_F(s_{N}^-) \delta t/2} \Big],
\nonumber\\
\label{eq:IFS}
\eea
\end{widetext}
where
\bea
K(s_{k+1}^{\pm},s_{k}^{\pm}) =\langle s_{k+1}^+|e^{-iH_S
\delta t}| s_{k}^+\rangle \langle s_{k}^- |e^{iH_S\delta t}|
s_{k+1}^-\rangle
\eea
is the propagator matrix for the subsystem. We have also used the
short notation $W$ for bath operators that are evaluated along the
path,
\bea
W_F(s)&= &H_F+\langle s| V_{SF}| s\rangle,
\nonumber\\
W_B(s)&=& H_B+\langle s| V_{SB}|s \rangle.
\eea
In the next sections we explain how we compute the bosonic and
fermionic IFs. The former has a closed analytic form in certain
situations. The latter is computed only numerically.

\subsection{Bosonic IF}
\label{Sec3b} We present the structure of the bosonic IF in two
separate models, corresponding to different types of subsystem-boson
bath interactions. In both cases the bosonic bath is prepared in a
canonical state of inverse temperature $\beta_{ph}=1/T_{ph}$,
\bea \rho_{B}=e^{-\beta_{ph}H_{B}}/{\rm Tr}_B[e^{-\beta_{ph}H_{B}}].
\eea
{\it Displacement interaction model},  $v_{p,p'}^B=0$ and
$\zeta_{p,p'}^B=0$. Given the remaining linear displacement-polarization
interaction, an analytic form for the bosonic IF can be written, the
so-called ``Feynman-Vernon" influence functional (FV IF) \cite{FV}.
In its time-discrete form, the bosonic IF is given by an exponent
with pairwise interactions along the path \cite{QUAPI1,*QUAPI2}
\bea I_B(s_0^{\pm},..., s_{N}^{\pm}) =
\exp\left[ -\sum_{k=0}^{N}\sum_{k'=0}^k
(s_k^+-s_k^-)(\eta_{k,k'}s_{k'}^+-\eta_{k,k'}^*s_{k'}^-)\right].
\label{eq:IBS}
\eea
The coefficients $\eta_{k,k'}$ are additive in the number of thermal
baths, and they depend on these baths' spectral functions and initial
temperatures \cite{QUAPI1,*QUAPI2}. For completeness, these coefficients are
included in Appendix A.

{\it Boson scattering model}, $\xi_{p}^B=0$. The bosonic IF can now
be computed numerically, by using the trace formula for bosons
\cite{Klich}
\bea {\rm Tr}_B[e^{M_1}
e^{M_2}...e^{M_k}]=\det[1-e^{m_1}e^{m_2}...e^{m_k}]^{-1}.
\label{eq:traceBS} \eea
Here $m_k$ is a single particle operator corresponding to a
quadratic bosonic operator
$M_k=\sum_{p,p'}(m_k)_{p,p'}b_p^{\dagger}b_{p'}$.  Application of
the trace formula to the bosonic IF (\ref{eq:IFS}) leads to
\bea
I_B&=& {\rm Tr}_B[e^{M_1}e^{M_2}...e^{M_k}\rho_B
]
\nonumber\\
&=& {\rm det} \Big \{  [\hat I_B+f_B]
- e^{m_1}e^{m_2}...e ^{m_k} f_B  \Big
\}^{-1}.
 \label{eq:levBS} \eea
The matrix $\hat I_{B}$ is an identity matrix, and  the function $f_B$
stands for the Bose-Einstein distribution,
$f_{B}=[e^{\beta_{ph}\omega}-1]^{-1}$. The determinant in Eq.
(\ref{eq:levBS}) can be evaluated numerically by taking into account
$L_B$ modes for the boson bath. This discretization implies a
numerical error. Generalizations, to include more that one bosonic
baths, are immediate.

\subsection{Fermionic IF}
\label{Sec3c}

The fermionic IF is computed numerically since an exact analytic
form is not known in the general strong coupling limit
\cite{MitraSpin1,MitraSpin2,SMarcus}. It is calculated by using the
trace formula for fermions \cite{Klich}
\bea {\rm Tr}_F[e^{M_1}
e^{M_2}...e^{M_k}]=\det[1+e^{m_1}e^{m_2}...e^{m_k}].
\label{eq:traceS} \eea
Here $m_k$ is a single particle operator corresponding to a
quadratic operator $M_k=\sum_{i,j}(m_k)_{i,j}c_i^{\dagger}c_j$.
In the next section we consider a model with two Fermi seas,
$H_F=H_L+H_R$, prepared in a factorized state of distinct grand
canonical states, $\rho_F=\rho_L\otimes\rho_R$, with
\bea \rho_{\nu}=e^{-\beta_{\nu}(H_{\nu}-\mu_{\nu}N_{\nu})}/{\rm
Tr}_F[e^{-\beta_{\nu}(H_{\nu}-\mu_{\nu}N_{\nu})}], \,\,\,\,\ \nu=L,R
\eea
Here $\beta_{\nu}=1/T_{\nu}$ stands for an inverse temperature, and
$\mu_{\nu}$ denotes the chemical potential of the $\nu$ bath.
Application of the trace formula to the fermionic IF in Eq.
(\ref{eq:IFS}) leads to
\bea I_F&=& {\rm Tr}_F[e^{M_1}e^{M_2}...e^{M_k}\rho_F ]
\nonumber\\
&=& {\rm det} \Big \{  [\hat I_L-f_L] \otimes  [\hat I_R-f_R]
+ e^{m_1}e^{m_2}...e ^{m_k} [ f_L \otimes f_R] \Big
\}.
\label{eq:levS} \eea
The matrices $\hat I_{\nu}$ are the identity matrices for the
$\nu=L,R$ space. The functions $f_L$ and $f_R$ are the bands
electrons' energy distribution,
$f_{\nu}=[e^{\beta_{\nu}(\epsilon-\mu_{\nu})}+1]^{-1}$. The
determinant in Eq. (\ref{eq:levS}) can be evaluated numerically by
taking into account $L_s$ electronic states for each metal. This
discretization implies a numerical error.

\subsection{The iterative scheme}
\label{Sec3d}

The dynamics described by Equation (\ref{eq:IBFSS}) includes
long-range interactions along the path, limiting brute force direct
numerical simulations to very short times. The iterative scheme,
developed in Ref. \cite{IF1,*IF2}, is based on the observation that
in standard nonequilibrium situations and at finite temperatures
bath correlations exponentially die \cite{SMarcus,eggerP}, thus the
IF can be truncated beyond a memory time $\tau_c=N_s \delta t$,
corresponding to the time where bath correlations sustain. Here
$N_s$ is an integer, $\delta t$ is the discretized time step, and
the correlation time $\tau_c$ is dictated by the bias and
temperature. Roughly, for a system under a potential bias
$\Delta\mu$ and a temperature $T$, $\tau_c\sim \max\{1/T,\,\,
1/\Delta\mu\}$ \cite{IF1,*IF2}.
By recursively breaking the IF to include terms only within
$\tau_c$, we reach the following (non-unique) structure for the
$\alpha=B,F$ influence functional,
\bea
&&I_{\alpha}(s_0^{\pm},s_1^{\pm},s_2^{\pm},...,s_N^{\pm})\approx
 I_{\alpha}(s_0^{\pm},s_1^{\pm},...,s_{N_s}^{\pm})
I_{\alpha}^{(N_s)}(s_1^{\pm},s_2^{\pm},...,s_{N_s+1}^{\pm})
I_{\alpha}^{(N_s)}(s_2^{\pm},s_3^{\pm},...,s_{N_s+2}^{\pm})
...
\nonumber\\
&&\times
I_{\alpha}^{(N_s)}(s_{N-N_s}^{\pm},s_{N-N_s+1}^{\pm},...,s_N^{\pm}), \label{eq:IF}
\eea
where we identify the ``truncated IF", $I_{\alpha}^{(N_s)}$, as the ratio between
two IFs, with the numerator calculated with an additional time step,
\bea I_{\alpha}^{(N_s)}(s_k,s_{k+1},...,s_{k+N_s})=
\frac{I_{\alpha}(s_k^{\pm},s_{k+1}^{\pm},...,s_{k+N_s}^{\pm})}{I_{\alpha}(s_{k}^{\pm},s_{k+1}^{\pm},...,s_{k+N_s-1}^{\pm})}.
\nonumber\\
\label{eq:IFtruncS} \eea
The truncated IF is the central object in our calculations. For
fermions, its numerator and denominator are separately computed
using Eq. (\ref{eq:levS}). The bosonic IF is similarly computed with
the help of Eq. (\ref{eq:levBS}) when $\xi_{p}^B=0$. In the
complementary case, $\zeta_{p,p'}^B=0$ and $v_{p,p'}^B=0$, the
truncated-bosonic IF has a closed analytic form: Using Eq.
(\ref{eq:IBS}) we find that it comprises only two-body interactions,
of $s_{k+N_s}$ with the preceding spins, down to $s_k$,
\bea
&&I_{B}^{(N_s)}(s_k,s_{k+1},...,s_{k+N_s})=
\exp\left[ -\sum_{k'=k}^{k+N_s}
(s_{k+N_s}^+-s_{k+N_s}^-)(\eta_{k+N_s,k'}s_{k'}^+-\eta_{k+N_s,k'}^*s_{k'}^-)\right].
\label{eq:IBtruncS}
\nonumber\\
\eea
Based on the decompositions (\ref{eq:IFtruncS}) and
(\ref{eq:IBtruncS}), we time-evolve Eq. (\ref{eq:IBFSS}) iteratively,
by defining a multi-time reduced density matrix $\tilde
\rho_S(s_{k},s_{k+1},..,s_{k+N_s-1})$. Its initial value is given by
\bea
\tilde \rho_S(s_0^{\pm},... ,s_{N_s}^{\pm})
=
I_S(s_0^{\pm},... ,s_{N_s}^{\pm})
I_B(s_0^{\pm},... ,s_{N_s}^{\pm}) I_F(s_0^{\pm},... ,s_{N_s}^{\pm}).
\eea
Its evolution is dictated by
\bea
&&
\tilde \rho_S(s_{k+1}^{\pm},...,s_{k+N_s}^{\pm})=
\sum_{s_{k}^{\pm}} \tilde\rho_S(s_{k}^{\pm},...,s_{k+N_s-1}^{\pm})
K(s_{k+N_s}^{\pm},s_{k+N_s-1}^{\pm})
\nonumber\\
&&\times I_{F}^{(N_s)}(s_{k}^{\pm},...,s_{k+N_s}^{\pm})
I_{B}^{(N_s)}(s_{k}^{\pm},...,s_{k+N_s}^{\pm}). \label{eq:prop2S}
\eea
The time-local ($t_k=k\delta t$) reduced density matrix, describing
the state of the subsystem at a certain time, is reached by summing
over all intermediate states,
\bea \rho_S(t_k)= \sum_ {s_{k-1}^{\pm} ...s_{k-N_s+1}^{\pm}} \tilde
\rho_S(s_{k-N_s+1}^{\pm},...,s_{k}^{\pm}). \label{eq:prop3} \eea
%
The bosonic and fermionic IFs may be (and often this is the case)
characterized by different memory time. Thus, in principle we could
truncate the fermionic IF to include $N_s^F$ terms, and the bosonic
IF to include $N_s^B$ elements. However, the efficiency of the
computation is dictated by the longest memory time, thus, for
convenience, we truncate both IFs using the largest value,
identified by $N_s$.

By construction, this iterative approach conserves the trace of the
reduced density matrix, ensuring the stability of
the iterative algorithm to long times \cite{QUAPI1,*QUAPI2}.
This property can be inferred from Eqs. (\ref{eq:IBFSS}) and (\ref{eq:IFS}), by using
the formal expressions for the truncated IFs, Eq. (\ref{eq:IF}) and
(\ref{eq:IFtruncS}).
To prover this property, we trace over the reduced density matrix at time $t$, identifying
$s_N=s_N^+=s_N^-$,
\bea
{\rm Tr}_S [\rho_S(t)]
&\equiv& \sum_{s_N}\langle s_N|\rho_S(t)|s_N\rangle
\nonumber\\
&=&
\sum_{{\bf s'^{\pm}}}
I_S({\bf s'^{\pm}})
I_F({\bf s'^{\pm}})
I_B({\bf s'^{\pm}})\delta(s_N^+-s_N^-)
\nonumber
\eea
%
Using the cyclic property of the trace, we note that both the
fermionic and bosonic IFs are independent of $s_N$, when
$s_N^+=s_N^-$. Therefore, the summation over the $s_N$ coordinate
reduces to a simple sum which can be performed using the
completeness relation for the subsystem states, resulting in
\bea
\sum_{s_N} \langle s_N|e^{-iH_S\delta t}|s_{N-1}^+\rangle
\langle s_{N-1}^-|e^{iH_S\delta t}|s_N\rangle
=
\delta (s_{N-1}^+-s_{N-1}^-).
\eea
%
Iterating in this manner we conclude that
\bea
{\rm Tr}_S [\rho_S(t)]
&\equiv& \sum_{s_N}\langle s_N|\rho_S(t)|s_N\rangle
\nonumber\\
&=&\sum_{{\bf s'^{\pm}}} I_S({\bf s'^{\pm}}) I_F({\bf s'^{\pm}})
I_B({\bf s'^{\pm}}) \delta(s_{N}^+-s_{N}^-)
\delta(s_{N-1}^+-s_{N-1}^-) ...  \delta(s_{1}^+-s_{1}^-)
\delta(s_{0}^+-s_{0}^-)
\nonumber\\
&=&\sum_{s_0}\langle s_0|\rho_S(0)|s_0\rangle ={\rm Tr}_S[\rho_S(0)]
\eea
We emphasize that the trace conservation is maintained even with the
use of the truncated form for the IFs. Moreover, it holds
irrespective of the details of the bath and the system-bath
interaction form. It is also obeyed in the more general case,  Eq.
(\ref{eq:IBFSP}). Equation (\ref{eq:prop2S}) [and its generalized
form, Eq. (\ref{eq:prop3S}) below], describe a linear map. Its fixed
points are stable if the eigenvalues of the map have modulus less
than one, which is the case here. 
Thus, our scheme is expected to approach a stationary-state in the long time limit.

\subsection{Expectation values for operators}
\label{Sec3e}

Besides the reduced density matrix, we can also acquire the time
evolution of several expectation values. Adopting the Hamiltonian
(\ref{eq:HS}), we illustrate next how we achieve the charge
current behavior. For simplicity, we consider the case with only two
fermionic reservoirs, $\nu=L,R$. The current operator, e.g., at the
$L$ bath is defined as the time derivative of the number operator.
The expectation value of this current is given by
\bea
j_L=
-\frac{d}{dt} {\rm Tr} [\rho N_L], \,\,\,\,\, N_L\equiv\sum_{j\in L}c_j^{\dagger}c_j
\eea
We consider the time evolution of the related exponential operator
$e^{\lambda N_L}$, with $\lambda$ a real number that is taken to
vanish at the end of the calculation,
\bea \langle  N_L(t) \rangle &\equiv& {\rm Tr} \left[\rho  N_L(t)\right]
\nonumber\\
&=&\lim_{\lambda \rightarrow 0} \frac{\partial}{\partial \lambda} {\rm
Tr}\big[\rho(0) e^{iHt}e^{\lambda  N_L}e^{-iHt} \big].
\label{eq:At} \eea
As before, the initial condition is factorized at $t=0,$
$\rho(0)=\rho_S(0)\otimes \rho_B\otimes \rho_F$. The trace is
performed over subsystem and reservoirs degrees of freedom. By
following the same steps as  in Eqs.
(\ref{eq:rhoSP})-(\ref{eq:pathP}), we reach the path-integral
expression
\begin{widetext}
\bea
&&\langle e^{\lambda N_L(t)}\rangle=
\sum_{s_0^{\pm}}\sum_{s_1^{\pm}}...\sum_{s_{N-1}^{\pm}}\sum_{s_N}
{\rm Tr}_B
{\rm Tr}_F  \Big[e^{\lambda  N_L}\langle s_N|\mathcal
G^{\dagger} |s^+_{N-1}\rangle \langle s_{N-1}^+| \mathcal
G^{\dagger} |s^+_{N-2}\rangle ...
\nonumber\\
&\times& \langle s_0^+| \rho(0)|s_0^{-}\rangle ...
 \langle s^-_{N-2}| \mathcal G  |s^-_{N-1}\rangle \langle
s_{N-1}^-|\mathcal G  |s_N\rangle \Big ]. \label{eq:PathS}
\eea
\end{widetext}
Factorizing the time evolution operators using Eq. (\ref{eq:trotter1P}),
we accomplish the compact form
\bea
\langle e^{\lambda  N_L(t)}\rangle&=&\sum_{\bf s'^{\pm}} I_S({\bf s'^{\pm}})
I_B({\bf s'^{\pm}})\tilde I_F({\bf s'^{\pm}})\delta(s_N^+-s_N^-).
\nonumber
\label{eq:NLS}
\eea
The terms $I_S$ and $I_B$ are given in Eq. (\ref{eq:IFS}). The
fermionic IF accommodates an additional exponent,
\bea
&&\tilde I_F({\bf s'^{\pm}})={\rm Tr}_F\Big[e^{\lambda N_L}
e^{-iW_F(s_N^+) \delta t/2} e^{-iW_F(s_{N-1}^+) \delta t}...
e^{-iW_F(s_{0}^+) \delta t/2} \rho_F e^{iW_F(s_0^-) \delta
t/2}....
e^{iW_F(s_{N-1}^-) \delta t} e^{iW_F(s_{N}^-) \delta t/2} \Big].
\nonumber
\label{eq:IFcurrentS}
\eea
We can time evolve the operator $\langle e^{\lambda N_L}\rangle$ by
using the iterative scheme of Sec. III.D, by truncating the bosonic
and fermionic IFs up to the memory time $\tau_c=N_s\delta t$, for
several values of $\lambda$. We then take the numerical derivative
with respect to $\lambda$ and $t$, to attain the charge current
itself.

The approach explained here could be used to explore several
fermionic operators, for example,
the averaged current $j_{av}=(j_L-j_R)/2$. The minus sign in front
of $j_R$ originates from the sign notation, with the current defined
positive when flowing $L$ to $R$. The implementation of a heat
current operator, describing the heat current flowing between two
bosonic reservoirs, requires first the derivation of an analytic
form for the bosonic IF, an expression analogous to the FV IF, and
the subsequent time discretization of this IF, to reach an
expression analogous to (\ref{eq:IBS}).

\subsection{Expression for multilevel subsystems and general interactions}
\label{Sec3f}

So far we have detailed the iterative time evolution scheme for the
spin-boson-fermion model (\ref{eq:HS}). The procedure can be
extended, to treat more complex cases. Based on the general
principles outlined in Sec. \ref{Sec3d}, one notes that the
path-integral expression (\ref{eq:IBFSP}) can be evaluated
iteratively by generalizing Eq. (\ref{eq:prop2S}) to the form
\bea &&\tilde \rho_S(v_{k+1}^{\pm},...,v_{k+N_s}^{\pm})=  \nonumber\\
&&
\sum_{v_{k}^{\pm}} \tilde\rho_S(v_{k}^{\pm},...,v_{k+N_s-1}^{\pm})
K(m_{k+N_s}^{\pm},n_{k+N_s}^{\pm})
I_{F}^{(N_S)}(s_{k}^{\pm}, f_{k}^{\pm}, g_{k}^{\pm},...,s_{k+N_s}^{\pm},  f_{k+N_s}^{\pm}, g_{k+N_s}^{\pm})
\nonumber\\
&&\times
I_{B}^{(N_s)}(f_{k}^{\pm},  g_{k}^{\pm}, m_{k}^{\pm}, n_{k}^{\pm},...,f_{k+N_s}^{\pm}, g_{k+N_s}^{\pm}, m_{k+N_s}^{\pm}, n_{k+N_s}^{\pm})
\label{eq:prop3S} \eea
where we compact several variables,
$v_k^{\pm}=\{s_k^{\pm},f_k^{\pm},g_k^{\pm},m_k^{\pm},n_k^{\pm}\}$.
It should be noted that in cases when the IF is time invariant, as
in the molecular electronics case discussed below, one needs to
evaluate $I_B^{(N_s)}$ and $I_F^{(N_s)}$ only once, then use the
saved array to time-evolve the auxiliary density matrix.

\begin{figure}[htbp]
\vspace{-1mm} {\hbox{\epsfxsize=60mm \hspace{10mm}\epsffile{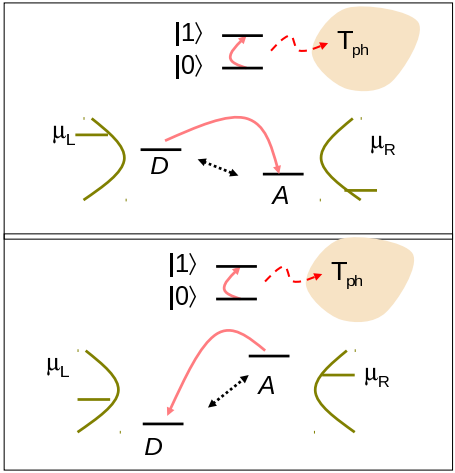}}}
\vspace{0mm} \caption{Molecular electronic rectifier setup. A biased
donor-acceptor electronic junction is coupled to an anharmonic mode,
represented by the two-state system with vibrational levels $|0\rangle$ and $|1\rangle$.
This molecular vibrational mode may further relax its energy to a
phononic thermal reservoir. 
This process is represented by a dashed arrow. Direct electron
tunneling element between D and A is depicted by a dotted double arrow. Top:
$\Delta \mu>0$. In our construction both molecular electronic levels
are placed within the bias window at large positive bias, resulting
in a large (resonant) current. Bottom: At negative bias the energy
of A is placed outside the bias window, thus the total charge current
is small.}
 \label{FigS}
\end{figure}

\section{Application: Molecular Rectifier}
\label{Sec4}

The functionality and stability of electron-conducting molecular
junctions are directly linked to heating and cooling effects
experienced by molecular vibrational modes in biased situations
\cite{NitzanVib,ExpVib1,ExpVib2,ExpVib3,McEuen,Selzer,Heating3,Heating2}.
In particular, junction heating and breakdown may occur once the
bias voltage exceeds typical molecular vibrational frequencies, when
the electronic levels are situated within the bias window, if energy
dissipation from the molecule to its environment is not efficient.

In this section
we study the dynamics and steady-state behavior of electrons and a
specific vibrational mode in a molecular conducting junction serving
as an electrical rectifier. The rectifier Hamiltonian is detailed in
Sec. \ref{Sec4a}. In Sec. \ref{Sec4b} we show that this model can be
mapped onto the spin-boson-fermion Hamiltonian (\ref{eq:HS}). This
allows us to employ the path-integral technique of Sec. \ref{Sec3}
for simulating the rectifier dynamics. The rectification mechanism
is explained in Sec. \ref{Sec4c}. Relevant expressions of a
(perturbative) Master equation method are described in Sec. \ref{Sec4Mas}, to
be compared to our path-integral based results in Sec. \ref{Sec4R}.
Convergence issues and computational aspects are discussed in Sec.
\ref{conv}.

\subsection{Rectifier Hamiltonian}
\label{Sec4a}

The D-A rectifier model includes a biased molecular electronic
junction and a selected (generally anharmonic) internal vibrational
mode which is coupled to an electronic transition in the junction
and to a secondary phonon bath, representing other molecular and
environmental degrees of freedom. In the present study we model the
anharmonic mode by a two-state system, and this model can already
capture the essence of the vibrational instability effect \cite{ET}.
For a schematic representation, see Fig. \ref{FigS}. This model
allows us to investigate the exchange of electronic energy with
molecular vibrational heating, and the competition between elastic
and inelastic transport mechanisms. Its close variant has been
adopted in Refs. \cite{Ora1,Ora2,Ora3} for studying the thermopower
and thermal transport of electrons in molecular junctions with
electron-phonon interactions, within the linear response regime.

We assume that the D molecular  group is strongly attached to the
neighboring $L$ metal surface, and that this unit is overall
characterized by the chemical potential $\mu_L$. Similarly, the A
group is connected to the metal $R$, characterized by $\mu_R$. At
time $t=0$ the D and A states are put into contact. Experimentally,
the $R$ metal may stand for an STM tip decorated by a molecular
group. This tip is approaching the D site which is attached to the
metal surface $L$. Once the D and A molecular groups are put into
contact, electrons can flow across the junction in two parallel
pathways: (i) through a direct D-A tunneling mechanism, and (ii)
inelastically, assisted by a vibration: excess electron energy goes
to excite the D-A vibrational
motion, and vice versa. 

The rectifier (rec) Hamiltonian includes the electronic Hamiltonian
$H_{el}$ with decoupled D and A states, the vibrational, two-state
subsystem $H_{vib}$, electronic-vibrational coupling $H_I$,
a free phonon Hamiltonian $H_{ph}$, and the coupling of this
secondary phonon bath to the selected vibration,
\bea \bar H_{rec}=H_{el} + H_{vib} + H_I+  H_{ph} + H_{vib-ph}.
\label{eq:HA} \eea
The electronic (fermionic) contribution $H_{el}$ attends for all
fermionic terms besides the direct D and A tunneling term, which for
convenience is included in $H_I$,
\bea H_{el}&=& H_{M}+H_L^0 +H_R^0+ H_{C}
\nonumber\\
H_M&=&\epsilon_dc_d^{\dagger}c_d + \epsilon_ac_a^{\dagger}c_a
\nonumber\\
H_{L}^0&=&\sum_{l\in L} \epsilon_{l}
c_l^{\dagger}c_l;\,\,\,\,\,\,\,
H_{R}^0=\sum_{r\in R} \epsilon_{r}
c_r^{\dagger}c_r.
\nonumber\\
H_C&=&
 \sum_{l} v_l\left(c_l^{\dagger}c_d +
c_d^{\dagger}c_l\right) + \sum_{r} v_r\left( c_r^{\dagger}c_a +
c_a^{\dagger}c_r\right).
\label{eq:HelA}
\eea
$H_M$ stands for the molecular electronic part including two
electronic states, a donor D and an acceptor A. $c_{d/a}^{\dagger}$
($c_{d/a}$) is a fermionic creation (annihilation) operator of an
electron on the D or A sites, of energies $\epsilon_{d,a}$. The two
metals, $H_{\nu}^0$, $\nu=L,R$, are each composed of a collection of
noninteracting electrons. The hybridization of the D state to the
left ($L$) bath, and similarly, the coupling of the A site to the
right ($R$) metal, are described by $H_C$.
The vibrational Hamiltonian includes a special nuclear anharmonic
vibrational mode of frequency $\omega_0$,
\bea H_{vib}= \frac{\omega_0}{2} \sigma_z.
\eea
The displacement of this mode from equilibrium is coupled to an
electron transition in the system, with an energy cost $\kappa$,
resulting in heating and/or cooling effects,
\bea
H_{I}= \left(\kappa\sigma_x + v_{da} \right)\left(c_d^{\dagger}c_a
+ c_a^{\dagger}c_d \right). \eea
Besides the electron-vibration coupling term, $H_I$ further includes
a direct electron tunneling element between the D and the A states,
of strength $v_{da}$. Electron transfer between the two metals can
therefore proceed through two mechanisms: coherent tunneling and
vibrational-assisted inelastic transport. 

The selected vibrational mode may couple to many other phonons,
either internal to the molecules or external, grouped into a harmonic
reservoir,
\bea
H_{ph}&=&\sum_{p}\omega_pb_p^{\dagger} b_p
\nonumber\\
H_{vib-ph}&=&\sigma_x\sum_p \xi_p^B\left( b_p^{\dagger}+b_p\right)
\eea
The Hamiltonian $H_{vib-ph}$ corresponds to a
displacement-displacement interaction type.

The motivation behind the choice of the two-level system (TLS) mode
is twofold. First, as we showed in Ref. \cite{ET}, the development
of vibrational instability in the D-A rectifier does not depend on
the mode harmonicity, at least in the weak electron-phonon coupling
limit. Since it is easier to simulate a truncated mode with our
approach, rather than a harmonic mode, we settle on the TLS model.
Second, while there are many studies where a perfectly harmonic mode
is assumed, for example, see Refs. \cite{Rabani,ThossExact,eggerP},
to the best of our knowledge our work is the first to explore
electron conduction in the limit of strong vibrational
anharmonicity.

\subsection{Mapping to the spin-boson-fermion model}
\label{Sec4b}

We diagonalize the electronic part of the Hamiltonian $H_{el}$
to acquire, separately, the exact eigenstates for the $L$-half and
$R$-half ends of $H_{el}$,
\bea H_{el}&=&H_L+H_R
\nonumber\\
H_L&=&\sum_{l} \epsilon_l a_{l}^{\dagger}a_l, \,\,\,
H_R=\sum_{r}\epsilon_r a_{r}^{\dagger}a_r. \eea
Assuming that the reservoirs are dense, their new operators are
assigned energies that are the same as those before diagonalization.
The D and A (new) energies are assumed to be placed within a band of
continuous states, excluding the existence of bound states. The old
operators are related to the new ones by \cite{Mahan}
\bea c_d&=&\sum_{l}\lambda_l a_l, \,\,\,\,\,\,\
c_l=\sum_{l'}\eta_{l,l'}a_{l'} \nonumber\\
c_a&=&\sum_{r}\lambda_r a_r, \,\,\,\,\,\,\
c_r=\sum_{r'}\eta_{r,r'}a_{r'},
 \eea
where the coefficients, e.g., for the $L$ set, are given by
\bea \lambda_l&=&\frac{v_l}{\epsilon_l-\epsilon_d-\sum_{l'}
\frac{v_{l'}^2}{\epsilon_l-\epsilon_{l'}+i\delta}} \nonumber\\
\eta_{l,l'}&=&\delta_{l,l'}-\frac{v_l\lambda_{l'}}{\epsilon_l-\epsilon_{l'}+i\delta}.
\label{eq:diagA}
 \eea
Similar expressions hold for the $R$ set. It is easy to derive the
following relation,
\bea
\sum_{l'}\frac{v_{l'}^2}{\epsilon_{l}-\epsilon_{l'}+i\delta}=
PP \sum_{l'}\frac{v_{l'}^2}{\epsilon_l-\epsilon_{l'}} -i \Gamma_L(\epsilon_l)/2,
\eea
with the hybridization strength ($v_j$ is assumed real),
%
\bea
\Gamma_L(\epsilon)=2\pi \sum_l v_l^2\delta(\epsilon-\epsilon_l).
\label{eq:GaA}
\eea
With the new operators, the Hamiltonian (\ref{eq:HA}) can be rewritten as
\bea  \bar H_{rec} &=& \sum_l  \epsilon_l a_l^{\dagger}a_l + \sum_r
 \epsilon_r a_r^{\dagger}a_r + \frac{\omega_0}{2} \sigma_z \nonumber\\
&+& \left(\kappa\sigma_x+v_{da}\right) \sum_{l,r} \left[\lambda_l^*
\lambda_r a_l^{\dagger} a_r + \lambda_r^* \lambda_l
a_r^{\dagger}a_l\right]
\nonumber\\
&+&
\sum_{p}\omega_pb_p^{\dagger} b_p +
\sigma_x\sum_p \xi_p^B\left( b_p^{\dagger}+b_p\right).
\label{eq:HTB} \eea
%
%
This Hamiltonian can be transformed into the spin-boson-fermion model
of zero energy spacing, using the unitary transformation
\bea
U^{\dagger}\sigma_z U=\sigma_x, \,\,\,\,\,  U^{\dagger}\sigma_x U=\sigma_z,
\eea
with $U=\frac{1}{\sqrt{2}}(\sigma_x+\sigma_z)$. The transformed
Hamiltonian $H_{rec}=U^{\dagger}\bar H_{rec}U$ includes a
$\sigma_z$-type electron-vibration coupling,
\bea  H_{rec} &=& \sum_l \epsilon_l a_l^{\dagger}a_l + \sum_r
\epsilon_r a_r^{\dagger}a_r+ \frac{\omega_0}{2} \sigma_x \nonumber\\
&+&
 \left(\kappa\sigma_z + v_{da}\right) \sum_{l,r}
\left[\lambda_l^* \lambda_r a_l^{\dagger} a_r + \lambda_r^*
\lambda_l a_r^{\dagger}a_l\right]
\nonumber\\
&+&
\sum_{p}\omega_pb_p^{\dagger} b_p +
\sigma_z\sum_p \xi_p^B\left( b_p^{\dagger}+b_p\right)
.
\label{eq:HH}
\eea
It describes a spin (TLS) coupled diagonally to two fermionic
environments and to a single boson bath. One can immediately confirm
that this Hamiltonian is accounted for by Eq. (\ref{eq:HS}). To
simplify our notation, we further identify the electronic-vibration
effective coupling parameter
\bea
\xi_{l,r}^F=\kappa\lambda_l^{*}\lambda_r.
\label{eq:FermC}
\eea
For later use we also define the spectral function of the secondary
phonon bath as
\bea J_{ph}(\omega)=\pi\sum_{p}(\xi^{B}_{p})^2\delta(\omega-\omega_p).
\label{eq:ohmic}
\eea
In our simulations below we adopt an ohmic function,
\bea
J_{ph}(\omega)=\frac{\pi K_d}{2}\omega e^{-\omega/\omega_c},
\label{eq:ohmic1}
\eea
with the dimensionless Kondo parameter $K_d$, characterizing
subsystem-bath coupling, and the cutoff frequency $\omega_c$.

As an initial condition for the reservoirs, we assume canonical
distributions with the boson-phonon bath distribution following
$\rho_{B}=e^{-\beta_{ph}H_{ph}}/{\rm Tr}_B[e^{-\beta_{ph}H_{ph}}]$
and the electronic-fermionic initial density matrix obeying
$\rho_F=\rho_L\otimes \rho_R$, with
$\rho_{\nu}=e^{-\beta_{\nu}(H_{\nu}-\mu_{\nu}N_{\nu})}/{\rm
Tr}_F[e^{-\beta_{\nu}(H_{\nu}-\mu_{\nu}N_{\nu})}]$, $\nu=L,R$. This
results in the expectation values of the exact eigenstates,
\bea \langle a_l^{\dagger}a_{l'}
\rangle=\delta_{l,l'}f_L(\epsilon_l),\,\,\,\,\, \langle
a_r^{\dagger}a_{r'} \rangle=\delta_{r,r'}f_R(\epsilon_r),
\eea
where $f_L(\epsilon)=[\exp(\beta_L(\epsilon-\mu_L))+1]^{-1}$ denotes
the Fermi distribution function. An analogous expression holds for
$f_R(\epsilon)$. The reservoirs temperatures are denoted by
$1/\beta_{\nu}$; the chemical potentials are $\mu_{\nu}$.


\begin{figure}[htbp]
\vspace{0mm} {\hbox{\epsfxsize=80mm \hspace{0mm}\epsffile{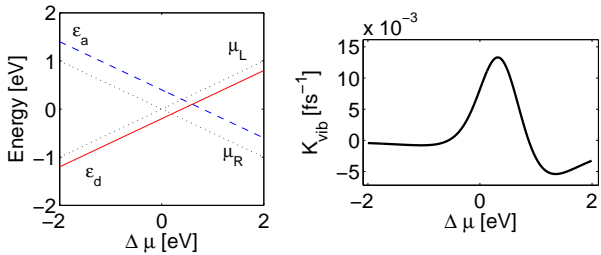}}}
\caption{Left panel: Energy of the donor (full line) and acceptor
states (dashed line). The dotted lines correspond to the chemical
potentials at the left and right sides. Right panel: Damping rate
$K_{vib}$. The junction's parameters are $\Gamma_{\nu}=1$,
$\beta_{\nu}=200$, $\kappa=0.1$, $\omega_0=0.2$, and
$\epsilon_d(\Delta\mu=0)=-0.2$, $\epsilon_a(\Delta\mu=0)=0.4$.
We used fermionic metals with a linear dispersion relations for
the original $H_{\nu}^0$ baths and  sharp cutoffs at $\pm 1$.
All energy parameters are given in units of eV. } \label{FigV}
\end{figure}

\begin{figure}[htbp]
\vspace{0mm} \hspace{0mm} {\hbox{ \hspace{10mm}\epsfxsize=120mm
\epsffile{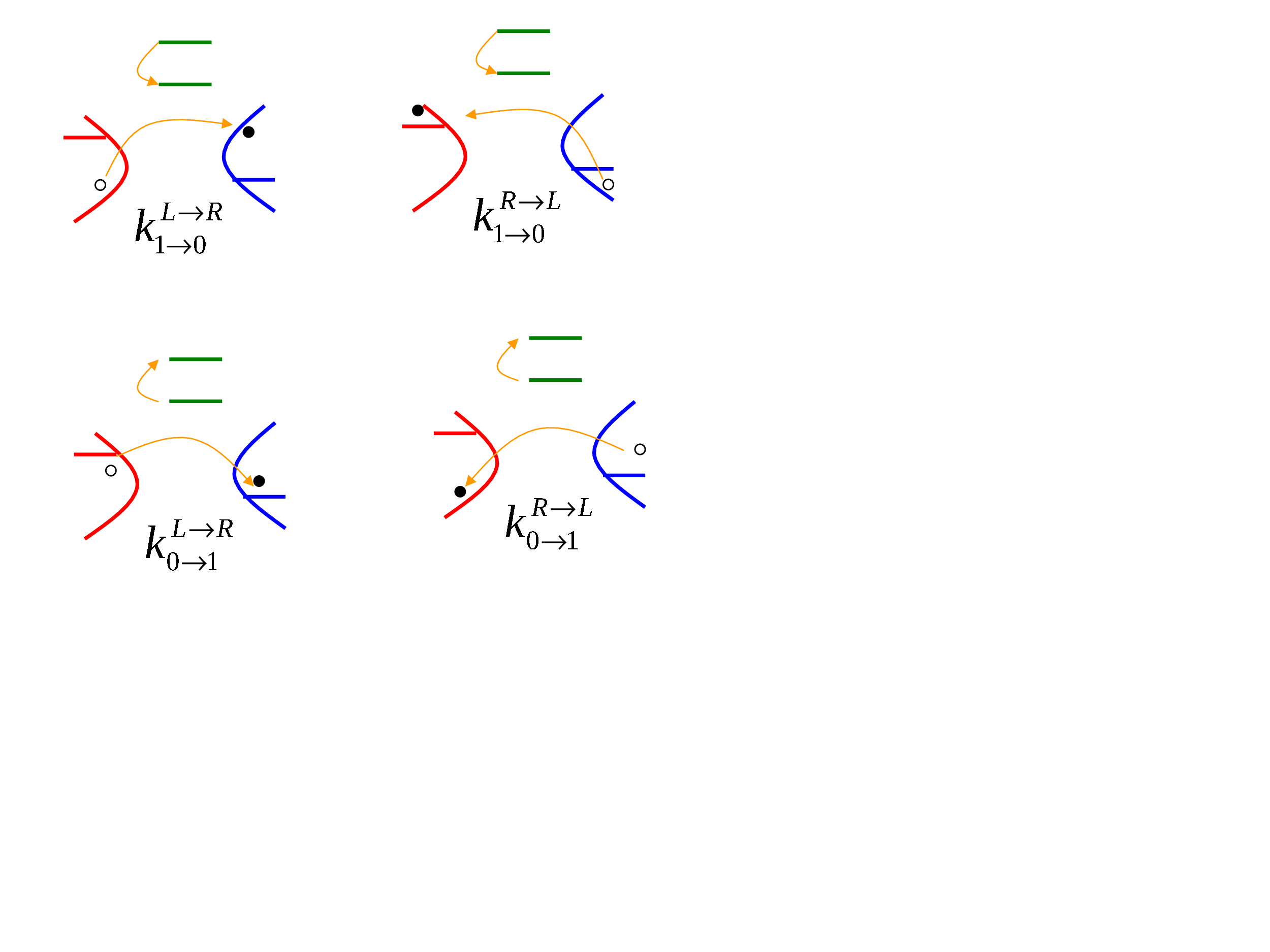}}} \vspace{-40mm}\caption{Scheme of the vibrational mode
excitation and relaxation processes.
A full circle represents an electron transferred; a hollow circle depicts
the hole that has been left behind.
} \label{FigR}
\end{figure}

\subsection{Rectifying mechanism}
\label{Sec4c}

We now explain the operation principles of the molecular rectifier.
In our construction the application of a bias voltage linearly
shifts the energies of the molecular electronic levels, D and A. In
equilibrium, we set $\epsilon_a<0$ and $\epsilon_d>0$. Under
positive bias, defined as $\mu_L-\mu_R>0$, the energy of the
acceptor level increases, and the donor level drops down, see Fig.
\ref{FigS}. When both levels are buried within the bias window, the
junction can support large currents. At negative bias the electronic
level A is positioned above the bias window, resulting in small
currents. For a scheme of the energy organization of the system, see
Fig. \ref{FigV} panel, left panel.

A generic mechanism leading to vibrational instabilities (and
eventually junction rupture) in D-A  molecular rectifiers has been
discussed in Ref. \cite{Lu}: At large positive bias, when the D
state is positioned above the acceptor level, electron-hole pair
excitations by the molecular vibration (TLS) dominate the mode
dynamics. This can be schematically seen in Fig. \ref{FigR}. The
second-order perturbation theory rate constant, to excite the
vibrational mode, while transferring an electron from $L$ to $R$,
$k_{0\rightarrow1}^{L\rightarrow R}$, overcomes other rates once the
density of states at the left end is positioned above the density of
states at the right side. This is the case at large positive bias,
given our construction. The rate $k_{0\rightarrow1}^{L\rightarrow
R}$ is defined next, in Sec. \ref{Sec4Mas}.


\subsection{Master equation ($v_{da}=0$)}
\label{Sec4Mas}

In the limit of weak electron-vibration coupling, once the direct
tunneling term is neglected, $v_{da}$=0, it can be shown that the
population of the truncated vibrational mode satisfies a kinetic
equation \cite{ET},
\bea &&\dot p_1=-\left(k_{1\rightarrow 0}^e +k_{1\rightarrow
0}^b\right) p_1 + \left( k_{0\rightarrow1}^e +  k_{0\rightarrow1}^b
\right)p_0,
\nonumber\\
&&p_0+p_1=1.
\label{eq:master}
\eea
The excitation ($k_{0\rightarrow 1}$) and relaxation
($k_{1\rightarrow 0}$) rate constants are given by a Fourier
transform of bath correlation functions of the operators  $F_e$ and $F_b$, defined as
\bea
F_e&=&\sum_{l,r} (\xi_{l,r}^F a_l^{\dagger}a_r +
\xi_{r,l}^Fa_r^{\dagger}a_l),
\nonumber\\
F_b&=&\sum_{p} \xi_{p}^B (b_{p}^{\dagger}+b_{p}), \eea
to yield
\bea k_{s\rightarrow s'}^e &=& \int_{-\infty}^{\infty}
 e^{i(\epsilon_s-\epsilon_{s'}) \tau} {\rm Tr}_F\left[\rho_F
 F_e(\tau) F_e(0)\right] d\tau
\nonumber\\
k_{s\rightarrow s'}^b &=& \int_{-\infty}^{\infty}
e^{i(\epsilon_s-\epsilon_{s'})\tau} {\rm Tr}_B\left[\rho_B F_b(\tau)
F_b(0)\right] d\tau. \label{eq:kTLS} \eea
Here $s=0,1$ and $\epsilon_1-\epsilon_0=\omega_0$.
The operators are given in the interaction representation, e.g.,
$a_l^{\dagger}(t)=e^{i H_L t}a_l^{\dagger}e^{-i H_Lt}$.

{\it Phonon-bath induced rates.} Expression (\ref{eq:kTLS})  can be simplified,
and the contribution of the phonon bath
to the vibrational rates reduces to
\bea k_{1\rightarrow0}^b&=&
\Gamma_{ph}(\omega_0)[f_{B}(\omega_0)+1],
\nonumber\\
k_{0\rightarrow1}^b&=&
 k_{1\rightarrow0}^b e^{-\omega_0\beta_{ph}},
\label{eq:bosonR}
\eea
where $f_{B}(\omega)=[e^{\beta_{ph}\omega}-1]^{-1}$ denotes the
Bose-Einstein distribution function. The damping rate is defined as
$\Gamma_{ph}(\omega)=2J_{ph}(\omega)$,
\bea \Gamma_{ph}(\omega)=2\pi\sum_{p}(\xi_{p}^B)^2
\delta(\omega_{p}-\omega). \label{eq:Gaph} \eea
For brevity, we ignore below the direct reference to frequency.

{\it Electronic-baths induced rates.} The electronic rate constants
(\ref{eq:kTLS}) include the following contributions \cite{ET},
\bea
k_{1\rightarrow 0}^e=k_{1\rightarrow 0}^{L\rightarrow
R}+k_{1\rightarrow 0}^{R\rightarrow L};\,\,\,\ k_{0\rightarrow
1}^e=k_{0\rightarrow 1}^{L\rightarrow R}+k_{0\rightarrow
1}^{R\rightarrow L},  \eea
satisfying
\bea
 k_{1\rightarrow0}^{L\rightarrow R}&=&
2\pi\kappa^2 
\sum_{l,r}|\lambda_l|^2|\lambda_r|^2f_L(\epsilon_l)(1-f_R(\epsilon_r))\delta(\omega_0+\epsilon_l-\epsilon_r)
\nonumber\\
 k_{0\rightarrow 1}^{L\rightarrow R}&=&
2\pi \kappa^2 
\sum_{l,r}|\lambda_l|^2|\lambda_r|^2f_L(\epsilon_l)(1-f_R(\epsilon_r))\delta(-\omega_0+\epsilon_l-\epsilon_r).
\label{eq:rate1} \eea
Similar relations hold for the right-to-left going excitations. The
energy in the Fermi function $f_{\nu}(\epsilon)$ is measured with
respect to the (equilibrium) Fermi energy, placed at
$(\mu_L+\mu_R)$, and we assume that the bias is applied
symmetrically, $\mu_L=-\mu_R$. The rates can be expressed in terms
of the fermionic $\nu=L,R$ spectral density functions
\bea J_{\nu}(\epsilon)&=&2\pi
\kappa\sum_{j\in\nu}|\lambda_{j}|^2\delta(\epsilon_{j}-\epsilon).
\eea
Using Eq. (\ref{eq:diagA}) we resolve this as a Lorentzian function,
centered around either the D or the A level,
\bea
J_L(\epsilon)&=&\kappa\frac{\Gamma_L(\epsilon)}{(\epsilon-\epsilon_d)^2+\Gamma_L(\epsilon)^2/4}
\nonumber\\ J_R(\epsilon)&=&
\kappa\frac{\Gamma_R(\epsilon)}{(\epsilon-\epsilon_a)^2+\Gamma_R(\epsilon)^2/4}.
\label{eq:spec}
\eea
The electronic hybridization $\Gamma_{\nu}(\epsilon)$ is given in
Eq. (\ref{eq:GaA}). Using these definitions, we express the
electronic rates [Eq. (\ref{eq:rate1})] by integrals ($s,s'$=0,1)
\bea
k_{s\rightarrow s'}^{\nu\rightarrow \nu'}&=&\frac{1}{2\pi}
\int_{-\infty}^{\infty}
f_{\nu}(\epsilon)\left[1-f_{\nu'}(\epsilon+(s-s')\omega_0)\right]
J_{\nu}(\epsilon)J_{\nu'}(\epsilon+(s-s')\omega_0)d\epsilon.
\label{eq:rate2}
\eea
%

{\it Observables.} Within this simple kinetic approach, junction
stability can be recognized by watching the TLS population in the
steady-state limit: population inversion reflects on vibrational
instability \cite{ET}. Solving Eq. (\ref{eq:master}) in the long
time limit we find that
\bea p_1=\frac{k_{0\rightarrow 1}^e+ k_{0\rightarrow 1}^b}
{k_{0\rightarrow 1}^e+k_{0\rightarrow 1}^b+k_{1\rightarrow 0}^e
+k_{1\rightarrow0}^b},\,\,\,\,\, p_0=1-p_1. \eea
A related measure is the {\it damping rate} $K_{vib}$ \cite{Lu}, depicted in Fig.
\ref{FigV} panel (b). It is defined as the difference between
relaxation and excitation rates,
\bea K_{vib}\equiv k_{1\rightarrow 0}^e+k_{1\rightarrow 0}^b
-\left(k_{0\rightarrow1}^e+k_{0\rightarrow 1}^b\right).
\label{eq:Kvib} \eea
Positive $K_{vib}$ indicates on a ``normal" thermal-like behavior,
when relaxation processes overcome excitations. In this case, the
junction remains stable in the sense that the population of the
ground state is larger than the population of the excited level. A
negative value for $K_{vib}$ evinces on the process of an {\it
uncontrolled} heating of the molecular mode, eventually leading to
vibrational instability and junction breakdown.

In the steady-state limit, the charge current $j$, flowing from $L$ to $R$, is given by \cite{ET}
\bea
j=p_1\left( k_{1\rightarrow 0}^{L\rightarrow R} -  k_{1\rightarrow 0}^{R\rightarrow L} \right)
+p_0 \left( k_{0\rightarrow 1}^{L\rightarrow R} -  k_{0\rightarrow 1}^{R\rightarrow L} \right).
\eea
This relation holds even when the TLS is coupled to an additional
boson bath. Note that in the long time limit the current that is
evaluated at the left end $j_L$ is equal to $j_R$. Therefore, we
simple denote the current by $j$ in that limit.

Master equation calculations proceed as follows. We set the
hybridization energy $\Gamma_{\nu}$ as an energy independent
parameter, and evaluate the fermionic spectral functions
$J_{\nu}(\epsilon)$ of Eq. (\ref{eq:spec}). With this at hand, we
integrate (numerically) Eq. (\ref{eq:rate2}), and gain the
fermionic-bath induced rates. The phonon bath-induced rates
(\ref{eq:bosonR}) are reached by setting the parameters of the
spectral function $J_{ph}$, to directly obtain $\Gamma_{ph}$, see
Eq. (\ref{eq:Gaph}). Using this set of parameters, we evaluate the
levels occupation and the charge current directly in the
steady-state limit. We can also time evolve the set of differential
equations (\ref{eq:master}), to obtain the trajectory $p_{1,0}(t)$.


\begin{figure}[htbp]
\vspace{0mm} {\hbox{\epsfxsize=70mm \hspace{0mm}\epsffile{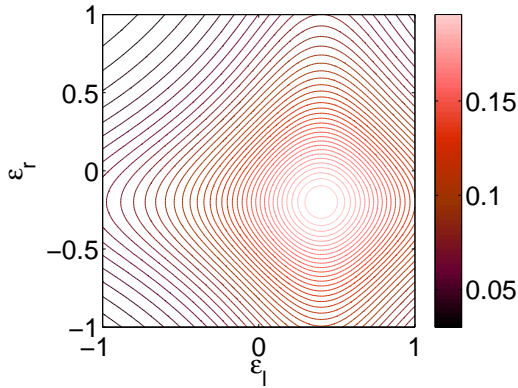}}}
\caption{Absolute value of the quantity $ \pi \rho \xi_{l,r}^F$.
The figure was generated by
discretizing the reservoirs, using bands extending from $-D$ to $D$,
 $D=1$, with $N_L=200$ states per each band a linear dispersion relation and a
constant density of states for the $H_{L,R}^0$ reservoirs, with a constant density
of states
$\rho=N_L/2D$. Electron-vibration coupling is given by $\kappa=0.1$.} \label{Figvlr}
\end{figure}


\subsection{Results}
\label{Sec4R}
We simulate the dynamics of the subsystem in the spin-boson-fermion
Hamiltonian (\ref{eq:HH}) using the path-integral approach of Sec.
\ref{Sec3}. In order to retrieve the vibrational mode occupation in
the original basis in which Eq. (\ref{eq:HTB}) is written, we rotate
the reduced density matrix $\rho_S(t)$ back to the original basis by
applying the transformation
$U=\frac{1}{\sqrt{2}}(\sigma_x+\sigma_z)$,
\bea
\bar \rho_S(t)=U\rho_S(t) U.
\eea
The diagonal elements of $\bar \rho_S(t)$, correspond to the
vibrational mode occupation, the ground state $|0\rangle$ and the
excited state $|1\rangle$,
\bea
p_0(t)=\langle 0 | \bar \rho_S(t)|0\rangle\,\,\,\,\,\, p_1(t)=\langle
1|\bar \rho_S(t)|1\rangle. \eea
As an initial condition we usually take
$\rho_S(0)=\frac{1}{2}(-\sigma_x+\hat I_s)$, $\hat I_s$ is a $2\times
2$ unit matrix. Under this choice, $\bar \rho_S(0)$ has only its
ground state populated.

Our simulations are performed with the following setup, displayed in
the left panel of Fig. \ref{FigV}:
In the absence of a bias voltage we assign the
donor the energy $\epsilon_d=-0.2$ and the acceptor the value
$\epsilon_a=0.4$. These molecular electronic states are assumed to
linearly follow the bias voltage. The right panel in Fig. \ref{FigV}
depicts the damping rate $K_{vib}$ in the absence of coupling to the
phonon bath, as evaluated using the Master equation method. This
measure becomes negative beyond $\Delta\mu\sim 0.85$, which
corresponds to the situation where the (bias shifted) donor energy
exceeds the acceptor by $\omega_0$, $\epsilon_d-\epsilon_a\gtrsim
\omega_0$; $\omega_0=0.2$. This results in a significant exchange of
electronic energy to heat, affecting junction's instability.

\subsubsection{Isolated mode}
\label{isol}

We study the time evolution of the vibrational mode occupation using
$v_{da}=0$ (unless otherwise stated), further decoupling it from a
secondary phonon bath, $K_d$=0.
%


{\it Electron-vibration interaction energy.} The interaction energy
of the subsystem (TLS) to the electronic degrees of freedom is
encapsulated in the matrix elements $\xi_{l,r}^F\equiv \kappa
\lambda_l^*\lambda_r$, see Eq. (\ref{eq:FermC}).
The strength of this interaction is measured by the dimensionless
parameter $\pi \rho(\epsilon_F) \xi_{l,r}^F$, which connects to the
phase shift experienced by Fermi sea electrons due to a scattering
potential, introduced here by the vibrational mode \cite{edge}.
Here, $\rho(\epsilon_F)$ stands for the density of states at the
Fermi energy. Using the parameters of Fig. \ref{FigV}, taking
$\kappa=0.1$, we show the absolute value of these matrix elements in
Fig. \ref{Figvlr}. The contour plot is mostly limited to values
smaller than 0.1, thus we conclude that this set of parameters
correspond to the weak coupling limit \cite{edge}. In this limit,
path-integral simulations should agree with Master equation
calculations, as we indeed confirm below. Deviations should be
expected at larger values, $\kappa\gtrsim 0.2$, and we study below
these cases.

{\it Units.} We perform the simulations in arbitrary units with
$\hbar\equiv 1$. One can scale all energies with respect to the
molecule-metal hybridization $\Gamma_{\nu}$. With $\Gamma_{\nu}=1$,
the weak coupling limit covers $\kappa/\Gamma_{\nu}\lesssim0.2$. To
present results in physical units, we assume that all energy parameters are
given in eV, and scale correspondingly the time unit and currents.

\begin{figure}[htbp]
\vspace{0mm} {\hbox{\epsfxsize=80mm \hspace{0mm}\epsffile{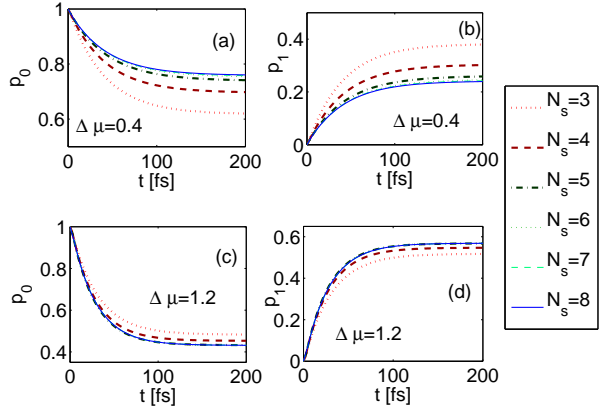}}}
\caption{Population dynamics and convergence behavior of the
truncated and isolated vibrational mode (TLS) with increasing $N_s$. (a)-(b)
Stable behavior at $\mu_L=-\mu_R=0.2$. (c)-(d) Population inversion
at $\mu_L=-\mu_R=0.6$. Other parameters are the same as in Fig.
\ref{FigV}. In all figures $\delta t$=1, $N_s=3$ (heavy dotted),
$N_s=4$ (heavy dashed), $N_s=5$ (dashed-dotted), $N_s=6$ (dotted),
$N_s=7$ (dashed) and $N_s=8$ (full). We used $L_s=30$ electronic
states at each fermionic bath with  sharp cutoffs at $\pm 1$.}
\label{FigC1}
\end{figure}

\begin{figure}[htbp]
\vspace{0mm} {\hbox{\epsfxsize=80mm \hspace{0mm}\epsffile{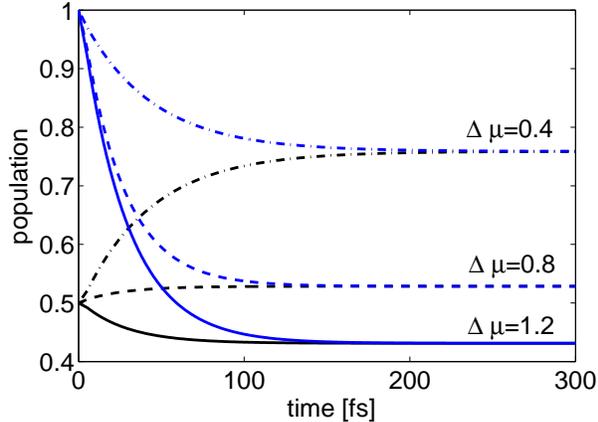}}}
\caption{
Independence of the population $p_0$ on the initial state
for different biases, $\Delta \mu=$ 0.4, 0.8, 1.2 top to bottom.
Other parameters are the same as in Fig. \ref{FigV} and Fig. \ref{FigC1}.
 } \label{incond}
\end{figure}

\begin{figure}[htbp]
\vspace{0mm} {\hbox{\epsfxsize=80mm \hspace{0mm}\epsffile{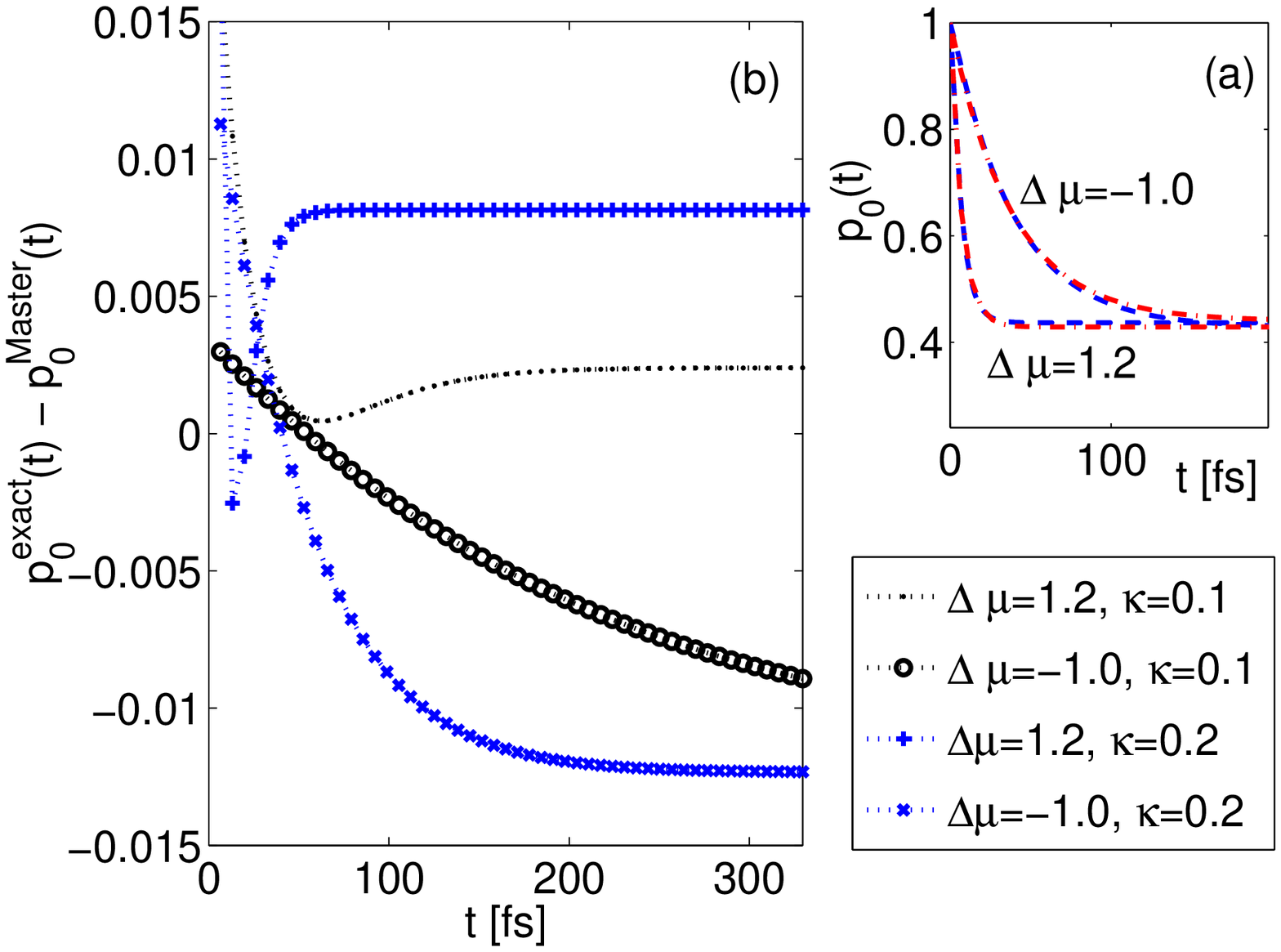}}}
\caption{Population dynamics, $p_0(t)$. (a) Comparison between exact
simulations (dashed) and Master equation results (dashed-dotted) at
$\kappa=0.2$. (b) Deviations between exact results and Master
equations for $\kappa=0.1$ (dot and $\circ$) and for $\kappa=0.2$
($+$ and $x$). Other parameters are as determined in  Fig.
\ref{FigV}.
 } \label{Figmast}
\end{figure}


{\it Dynamics.} We first focus on two representative values for the
bias voltage: In the low-positive bias limit a stable operation is
expected, reflected by a normal population, $p_0>p_1$. At large
positive bias population inversion may take place, indicating on the
onset of instability and potential junction rupture \cite{ET}.

Fig. \ref{FigC1} displays the TLS dynamics, and we present data for
different memory sizes $N_s\delta t$. At small positive bias,
$\epsilon_d-\epsilon_a<\omega_0$, the mode occupation is ``normal",
$p_0>p_1$. In particular, in panels (a)-(b) we discern the case
$\mu_L=-\mu_R=0.2$, resulting in the (shifted) electronic energies
$\epsilon_d=0$ and $\epsilon_a=0.2$. In this case the (converged)
asymptotic long-time population (representing steady-state values),
are $p_0^{ss}=0.76$ and $p_1^{ss}=0.24$. In contrast, when the bias
is large, $\mu_L=-\mu_R=0.6$, the electronic levels are shifted to
$\epsilon_d=0.4$ and $\epsilon_a=-0.2$, and electrons crossing the
junction discard their excess energy into the vibrational mode.
Indeed, we see in Fig. \ref{FigC1}(c)-(d) the process of {\it
population inversion}, $p_0^{ss}=0.43$ and $p_1^{ss}=0.57$. The TLS
approaches the steady-state value around $t_{ss}\sim 0.1$ ps.
Regarding convergence behavior, we note that at large bias
convergence is reached with a shorter memory size, compared to the
small bias case, as expected \cite{IF1}.

Fig. \ref{incond} exhibits the dynamics with different initial
conditions, demonstrating that the steady-state value is identical,
yet the timescale to reach the stationary limit may depend on the
initial state.

We compare the exact dynamics to the Master equation time evolution
behavior, reached by solving  Eq. (\ref{eq:master}). Panel (a) in
Fig. \ref{Figmast}  demonstrates excellent agreement for
$\kappa=0.2$, for both positive and negative biases. Below we show
that at this value Master equation's predictions for the charge
current deviate from the exact result. Panel (b) in Fig.
\ref{Figmast} focuses on the departure of Master equation data from
the exact values. These deviations are small, but their dynamics
indicate on the existence of high order excitation and relaxation
rates, beyond the second order rates of Sec. \ref{Sec4Mas}.

\begin{figure}[htbp]
\vspace{0mm} {\hbox{\epsfxsize=70mm \hspace{0mm}\epsffile{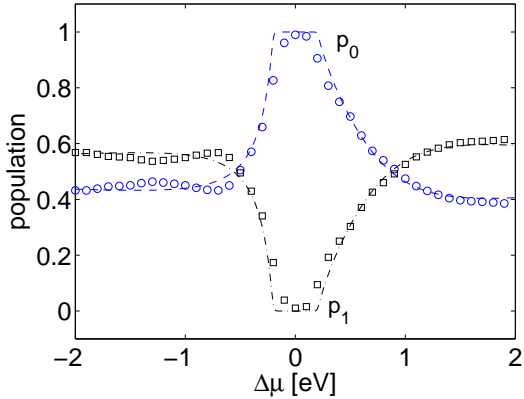}}}
\caption{Converged data for the population of the isolated vibrational mode
in the steady-state limit with $\kappa=0.1$. Other parameters are
the same as in Fig. \ref{FigV}. We display path-integral data for
$p_0$ ($\circ$) and $p_1$ ($\square$). Master equation results
appear as dashed line for $p_0$ and dashed-dotted line for $p_1$.
 } \label{Figbias1}
\end{figure}

\begin{figure}[htbp]
\vspace{0mm} {\hbox{\epsfxsize=70mm \hspace{0mm}\epsffile{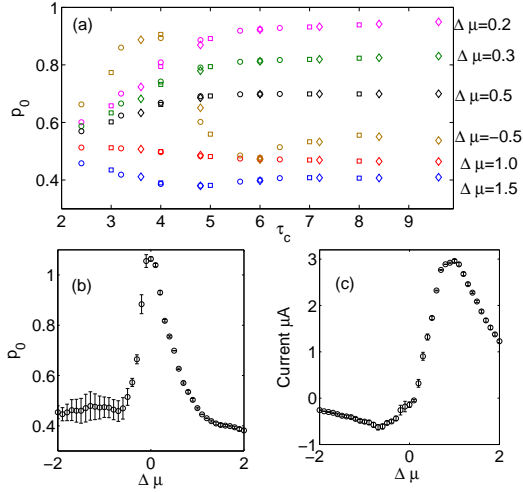}}}
\caption{(a) Convergence behavior of the population $p_0$ in the
steady-state limit for $\kappa=0.1$. Other parameters are the same
as in Fig. \ref{FigV}. Plotted are the steady-state values using
different time steps, $\delta t=0.8$ ($\circ$), $\delta t=1.0$
($\square$), and $\delta t=1.2$ ($\diamond$) at different biases, as
indicated at the right end. (b) Population mean and its standard deviation, utilizing
the last six points from panel (a). (c) Current mean and its standard deviation,
similarly attained from the data in panel (a).
 } \label{Conv1}
\end{figure}

{\it Steady-state characteristics.} The full bias scan of the steady-state
population is displayed in Fig. \ref{Figbias1}, and we compare
path-integral results with Master equation calculations, revealing
an excellent agreement in this weak coupling limit ($\kappa=0.1$).
The convergence behavior is presented in Fig. \ref{Conv1}, and we
plot the steady-state values as a function of memory size ($\tau_c$)
for three different time steps, for representative biases. The path-integral
results well converge at intermediate-to-large positive biases,
$\Delta \mu\gtrsim 0.2$. We had difficulty converging our results in
two domains: (i) At small-positive potential bias, $\Delta \mu
<0.2$. Here, large memory size should be used for reaching full
convergence;  decorrelation time approximately scales with
$1/\Delta \mu$. (ii) At large negative biases, $\Delta \mu< -0.4$
the current is very small as we show immediately. This implies poor
convergence at the range of $\tau_c$ employed. At these negative
biases the data oscillates with $\tau_c$, thus at negative bias it
is the {\it averaged} value for several-large $\tau_c$ which is
plotted in Fig. \ref{Figbias1}.

\begin{figure}[htbp]
\vspace{0mm} {\hbox{\epsfxsize=70mm \hspace{0mm}\epsffile{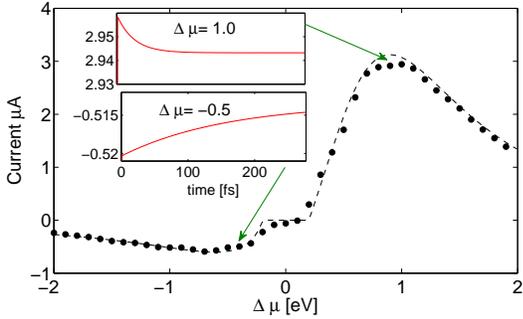}}}
\caption{Charge current in the steady-state limit for $\kappa=0.1$,
$K_d=0$.
Other parameters are the same as in Fig. \ref{FigV}.
Path-integral data ($\circ$), Master equation results (dashed). The
insets display transient results at $\Delta \mu=1.0$ eV (top) and
$\Delta \mu=-0.5$ eV (bottom).
 } \label{Figbias2}
\end{figure}

{\it Charge current.} We show the current characteristics in Fig.
\ref{Figbias2}, and confirm that the junction acts as a charge
rectifier. 
The insets display transient data, affirming that at large bias
steady-state is reached faster than in the low bias case.

\begin{figure}[htbp]
\vspace{0mm} {\hbox{\epsfxsize=80mm \hspace{0mm}\epsffile{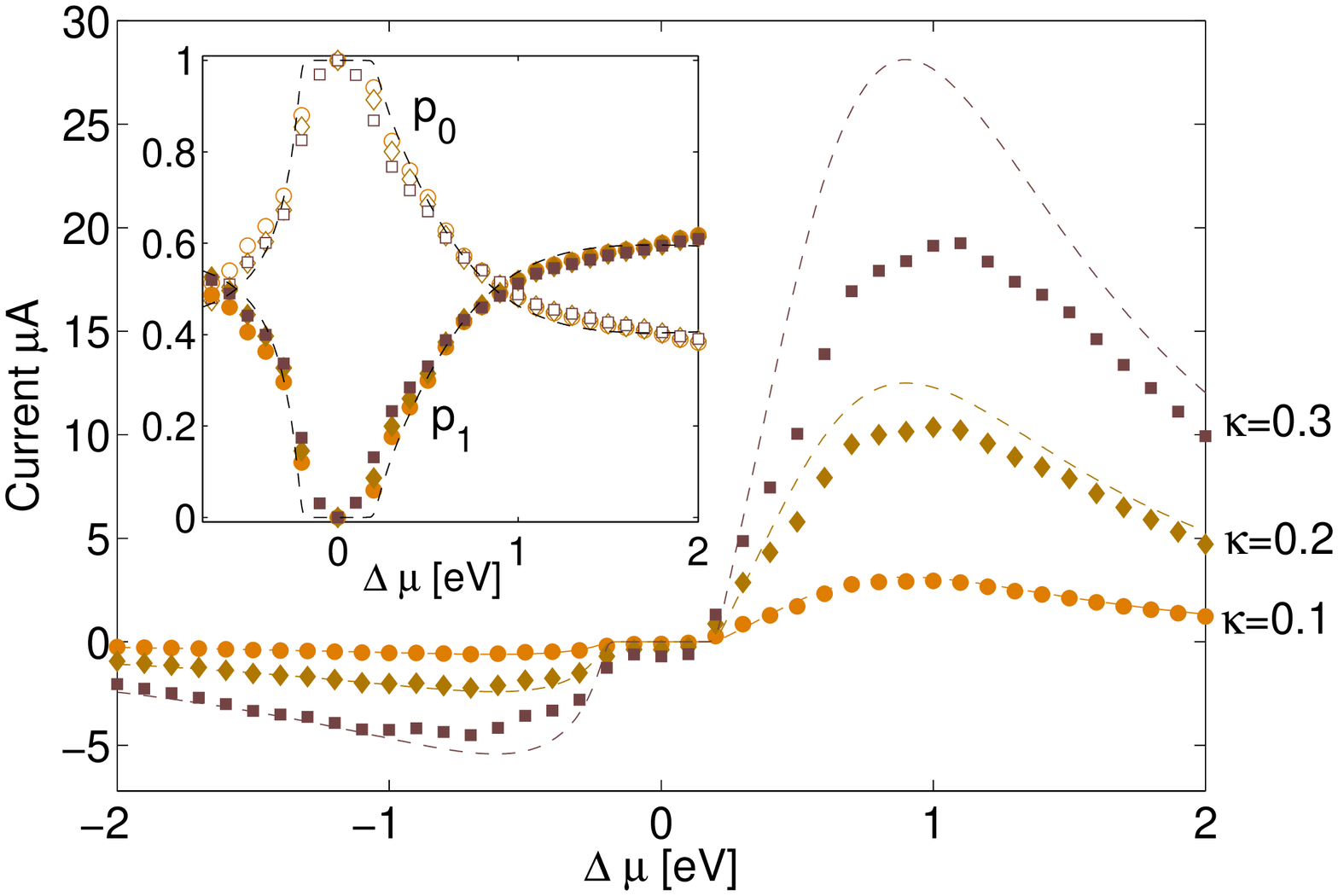}}}
\caption{Charge current and vibrational occupation in the
steady-state limit at different electron-vibration coupling.
Path-integral data is marked by symbols, $\kappa=0.1$ ($\circ$),
$\kappa=0.2$ ($\diamond$) and $\kappa=0.3$ ($\square$). The
corresponding Master equation results appear as dashed lines. Inset:
The population behavior in the steady-state limit for the three
cases $\kappa=0.1$ ($\circ$), $\kappa=0.2$ ($\diamond$) and
$\kappa=0.3$ ($\square$), with empty symbols for $p_0$ and filled
ones for $p_1$. Other parameters are the same as in Fig. \ref{FigV}.
 } \label{Figbias3}
\end{figure}

{\it Strong coupling.} Results at weak-to-strong couplings are shown
in Fig. \ref{Figbias3}. The value of the current, as reached from Master
equation calculations, scale with $\kappa^2$. In contrast, exact
simulations indicate that the current grows more slowly with $\kappa$, and it
displays clear deviations (up to $50\%$) from the perturbative
Master equation result at $\kappa=0.3$. Interestingly, the
vibrational occupation (inset) shows little sensitivity to the
coupling strength, and even at $\kappa=0.3$ the Master equation
technique provides an excellent estimation for the levels occupation. This
could be reasoned by the fact that excited levels occupation is
given by ratio of excitation rates to the sum of excitation and
relaxation rates. Such a ratio is (apparently) only weakly sensitive
to the value of $\kappa$ itself, even when high-order processes do
contribute to the current.


\begin{figure}[htbp]
\vspace{0mm} {\hbox{\epsfxsize=80mm \hspace{0mm}\epsffile{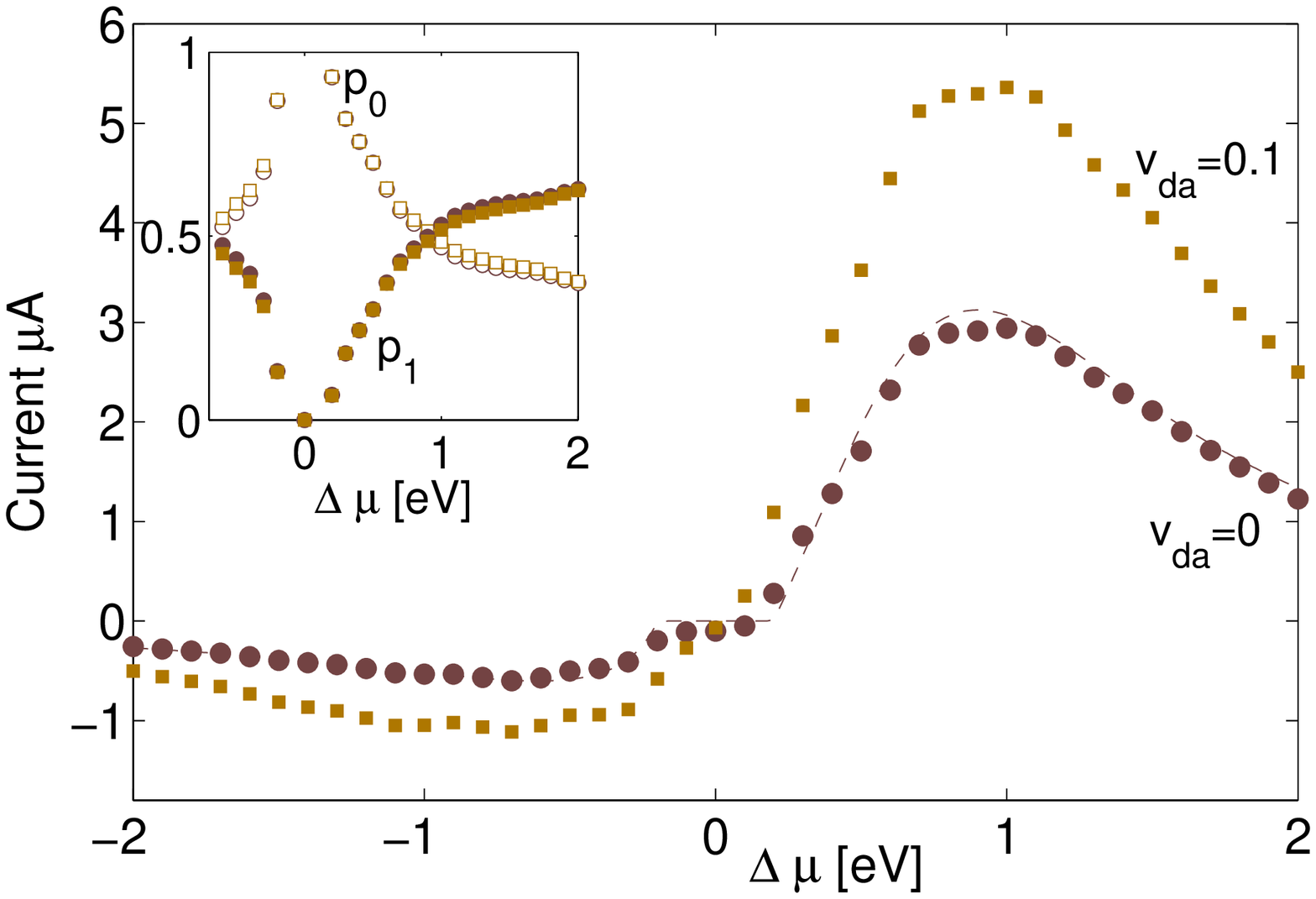}}}
\caption{Study of the contribution of different transport
mechanisms. $v_{da}=0$ ($\circ$), with Master equation results noted
by the dashed line, and $v_{da}=0.1$ ($\square$). The main plot
displays the charge current. The inset presents the vibrational
levels occupation, with empty symbols for $p_0$ and filled symbols
for $p_1$. Other parameters are the same as in Fig. \ref{FigV},
particularly, the vibrational-electronic coupling is $\kappa=0.1$. }
\label{Fig:Vda}
\end{figure}

{\it Direct tunneling vs. vibrational assisted transport.} Until
this point (and beyond this subsection) we have taken $v_{da}=0$. We
now evaluate the contribution of different transport mechanisms by
adding a direct D-A tunneling term, $v_{da}\neq0$ to our model
Hamiltonian. Electrons can now either cross the junction in a
coherent manner, or inelastically, by exciting/de-exciting the
vibrational mode. Fig. \ref{Fig:Vda} demonstrates that when the
vibration assisted transport energy $\kappa$ is identical in
strength to the direct tunneling element $v_{da}$, the overall
current is enhanced by about a factor of two, compared to the case
when only vibrational-assisted processes are allowed. We also note
that the occupation of the vibrational mode is barely affected by the
opening of the new electron transmission route (deviations are
within the convergence error). While we compare IF data to Master
equation results when $v_{da}=0$, in the general case of a nonzero
D-A tunneling term perturbative methods are more involved, and
techniques similar to those developed for the AH model should be
used \cite{NitzanVib,Millis,Galperin1,Galperin2,Jonas,
Wege,Thoss1,Thoss2}.

\subsubsection{Equilibration with a secondary phonon bath}
\label{thermal}

We couple the isolated-truncated vibrational mode to a secondary
phonon bath, and follow the mode equilibration with this bath and
the removal of the vibrational instability effect, as we increase
the vibrational mode-phonon bath coupling. As an initial condition,
the boson bath is assumed to be thermal with an inverse temperature
$\beta_{ph}$. This bath is characterized by an ohmic spectral
function (\ref{eq:ohmic1}) with the dimensionless Kondo parameter
$K_d$, characterizing subsystem-bath coupling, and the cutoff
frequency $\omega_c$.

\begin{figure}[htbp]
\vspace{0mm} {\hbox{\epsfxsize=80mm \hspace{0mm}\epsffile{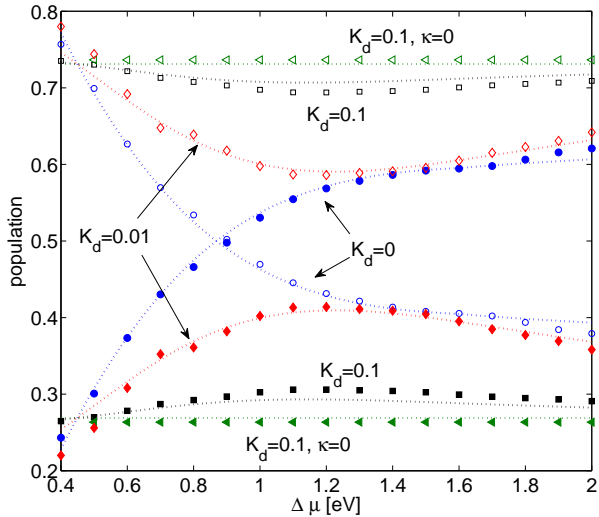}}}
\caption{Equilibration of the molecular vibrational mode with
increasing coupling to a secondary phonon bath. Path-integral
results, (full symbols for $p_1$, and empty symbols for $p_0$) with
$K_d=0$ ($\circ$), $K_d=0.01$ ($\diamond$), $K_d=0.1$ ($\square$),
and, $K_d=0.1$, $\kappa=0$ ($\triangleleft$). Unless otherwise
specified, $\kappa=0.1$, $\beta_{ph}=5$ and the spectral function
follows (\ref{eq:ohmic1}) with $\omega_c$=15. All other electronic
parameters are the same as in Fig. \ref{FigV}. Master equation
results appear in dotted lines.} \label{Fig:Ph1}
\end{figure}

{\it Population behavior.} We follow the mode dynamics to the
steady-state limit using the path-integral approach of Sec.
\ref{Sec3}. The bosonic IF is given in the appendix. We compare
exact results with Master equation predictions, and Fig.
\ref{Fig:Ph1} depicts our simulations. The following observations
can be made: (i) The vibrational instability effect is removed
already for $K_d=0.01$, though nonequilibrium effects are still
largely visible in the mode occupation. (ii) The vibrational mode is closed
to be equilibrated with the phonon bath once $K_d\sim 0.1$. (iii)
For the present range of parameters (large $\omega_c$, weak
subsystem-bath couplings), Master equation tools reproduce the
behavior of the vibrational mode.

\begin{figure}[htbp]
\vspace{0mm} {\hbox{\epsfxsize=80mm \hspace{0mm}\epsffile{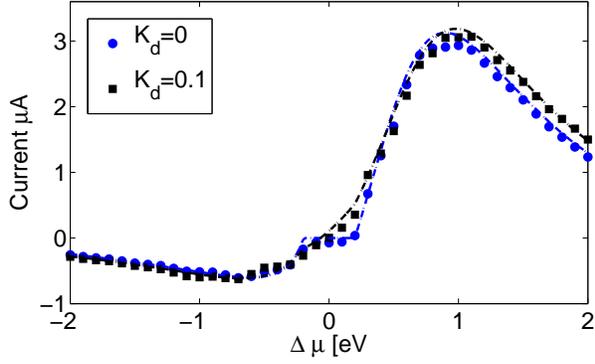}}}
\caption{Charge current for an isolated mode,$K_d=0$ ($\circ$), and
an equilibrated mode, $K_d=0.1$, $\beta_{ph}$=5, $\omega_c$=15
($\square$). Other electronic parameters are given in Fig.
\ref{FigV}. Master equation results appear in dashed-dotted lines.}
\label{Fig:Ph2}
\end{figure}

{\it Charge Current.} The role of the secondary phonon bath on the
charge current characteristics is displayed in Fig. \ref{Fig:Ph2}.
There are two main effects related to the presence of the phonon
bath: The step structure about zero bias is flattened when $K_d\sim
0.1$, and the current-voltage characteristics as a whole is slightly
enhanced at finite $K_d$, at large bias. Both of these effects are
excellently reproduced with the Master equation, and we conclude
that in this weak-coupling regime the presence of the phonon bath
does not affect the rectifying behavior of the junction. We have
also verified (not shown) that at stronger coupling, $\kappa=0.2$
(where Master equation fails), the thermal bath similarly affects
the current-voltage behavior.

An important observation is that the current itself does not testify
on the state of the vibrational mode, whether it is in a stable or
an unstable nonequilibrium state, and whether it is thermalized. The
study of the current characteristics itself ($j$ vs. $\Delta\mu$) is
therefore insufficient to determine junction stability. More
detailed information can be gained from the structure of the first
derivative, $dj/d(\Delta\mu)$, the local density of states, and the
second derivative, $d^2j/d\Delta\mu^2$, providing spectral features
\cite{IETS1,IETS2,IETS3}. In order to examine these quantities, our
simulations should be performed with many more bath states, to
eliminate possible spurious oscillations in the current (of small
amplitudes) that may result from the finite discretization of the
fermi baths.




\subsection{Convergence and Computational aspects}
\label{conv}

Convergence of the path-integral method should be verified with
respect to three numerical parameters: the number of states used to
mimic a fermi sea, $L_s$, the time step adopted, $\delta t$, and the
memory time accounted for, $\tau_c$. (i) {\it Fermi sea
discretization.} We have found that excellent convergence is
achieved for relatively ``small" fermi reservoirs, taking into
account $L_s>20$ states for each reservoir. In our simulations we
practically adopted $L_s=30$ for each Fermi bath. (ii) {\it
Time-step discretization.} The first criteria in selecting the value
of the time step $\delta t$ is that dynamical features of the
isolated vibrational mode should be observed. Using $\omega_0=0.2$,
the period of the bath-free Rabbi oscillation is $2
\pi/(\omega_0)\sim 30$, thus a time step of $\delta t\sim 1$ can
capture the details of the TLS oscillation. This consideration
serves as an ``upper bound" criteria. The second consideration
connects to the time discretization error, originates from the
approximate splitting of the total time evolution operator into a
product of terms, see Eq. (\ref{eq:trotter1P}). For the particular
Trotter decomposition employed, the leading error grows with $\delta
t^3 \times \left( [H_S,[V,H_S]]/12 +[V,[V,H_S]]/24 \right)$
\cite{Tannor} where $V=V_{SB}+V_{SF}+H_{B}+H_F$. The decomposition
is exact when the coupling of the subsystem to the reservoirs is
weak and the time-step is small, $\delta t\rightarrow 0$. For large
coupling one should take a sufficiently small time-step in order to
avoid significant error buildup. In the preset work, the
dimensionless coupling to the fermi sea $\pi\rho \kappa
\lambda_l^*\lambda_r$ is typically maintained lower that 0.3; the
dimensionless coupling to the boson bath is taken as $K_d=0.1$. The
value of $\delta t=0.6-1.2$ is thus sufficiently small for our
simulations. (iii) {\it Memory error}. Our approach assumes that
bath correlations exponentially decay resulting from the finite
temperature and the nonequilibrium condition. Based on this
assumption, the total influence functional was truncated to include
only a finite number of time steps $N_s$, where $\tau_c=N_s \delta
t$. The total IF is retrieved by taking the limit $N_s \rightarrow
N$, ($N=t/\delta t$). Our simulations were performed for
$N_s=3...9$, covering memory time up to $\tau_c=N_s\delta t\sim 10$.
The results displayed converged for $N_s\sim 7-9$ for $\delta t=1$.

Computational efforts can be partitioned into two parts: In the
initialization step the (time invariant) IFs are computed. The size
of the fermionic IF is $d^{2N_s}$, where $d$ is the dimensionality
of the subsystem (two in our simulations). The power of two at the
exponent results from the forward and backward time evolution
operators in the path-integral expression. This initialization
effort thus scales exponentially with the memory size accounted for.
The preparation of the bosonic IF is more efficient if the FV IF is
used \cite{QUAPI1,*QUAPI2}. In the second, time evolution, stage, we
iteratively apply the linear map (\ref{eq:prop2S}) or
(\ref{eq:prop3S}), a multiplication of two objects of length
$d^{2N_s}$. This operation linearly scales in time.

We now comment on the simulation time of a convergence analysis as
presented in Fig. \ref{Conv1}, covering three different time steps
and $N_s=3,...,9$. The MATLAB implementation of the computational
algorithm took advantage of the MATLAB built-in multi-threaded
parallel features and utilized 100$\%$ of all available CPU cores on
a node. When executed on one cluster node with two quad-core 2.2GHz
AMD Opteron cpus and 16GB memory, convergence analysis for of the
full voltage scan took about 4x24 hours and 250MB of memory.
Computations performed on the GPC supercomputer at the SciNet HPC
Consortium \cite{SciNet} were three times faster. Computational time
scales linearly with the simulated time $t$. For a fixed $N_s$
value, the computational effort does not depend on the system
temperature and other parameters employed.

\section{Summary}
\label{Sec5}

We have developed an iterative numerically-exact path-integral
scheme that can follow the dynamics, to the steady-state limit, of
subsystems coupled to multiple bosonic and fermionic reservoirs in
an out-of-equilibrium initial state. The method is based on the
truncation of time correlations in the influence functional, beyond
the memory time dictated by temperature and chemical biases. It
combines two techniques: the QUAPI method \cite{QUAPI1,*QUAPI2}, for
treating the dynamics of subsystems coupled to harmonic baths, and
the INFPI approach \cite{IF1,*IF2}, useful for following the evolution of a
subsystem when interacting with fermionic baths.

The method is stable, efficient, and flexible, and it allows one to
achieve transient and steady-state data for both the
reduced density matrix of the subsystem and expectation values of
operators, such as the charge current and energy current. The method
can be viewed as an extension of QUAPI, to incorporate fermions in
the dynamics. It could be further expanded to include time-dependent
Hamiltonians, e.g., pulsed fields.

To demonstrate the method usability in the field of molecular
conduction, we have applied the general scheme, and studied
vibrational dynamics in a molecular rectifier setup, where
vibrational equilibration with an additional phonon bath is allowed.
Our main conclusions in this study are the following: (i) The
vibrational instability effect disappears once the vibrational mode
is weakly coupled ($K_d\sim 0.01$) to an additional phonon bath that
can dissipate the excess energy. (ii) When $K_d\sim 0.1$, the
vibrational mode is equilibrated with the secondary phonon bath.
(iii) The charge current does not testify on vibrational heating and
instability. While we have performed those simulations using a
truncated vibrational mode, a TLS, representing an anharmonic mode,
we argue that the main characteristics of the vibrational
instability effect remain intact when the selected mode is made
harmonic \cite{ET}.

Our simulations indicate that Master equation methods can excellently
reproduce exact results at weak coupling, in the markovian limit.
More significantly, Master equation tools can be used beyond the
weak coupling limit ($\kappa\sim 0.3$), if only a qualitative
understanding of the junction behavior is inquired.
One should note that our Master equation technique treats the D and A coupling to the metals
{\it exactly}. It is perturbative only in the interaction of the vibrational mode to
the electrons, and to other phonon degrees of freedom.
In the case where tunneling transmission competes with phonon-assisted
transport, only path-integral simulations were provided, as more
involved Master equation methodologies should be developed in this case.

Our future objectives are twofold: (i) to improve the time-evolution
algorithm, and (ii) to employ the method for the study of other
problems in molecular electronics and phononics. By improving the
methodology, we would like to extend the usability of our method to
difficult parameter regimes (strong coupling), e.g., by developing
an equation-of-motion for the memory function \cite{Golosov,Cohen}.
This will also allow us to simulate more feasibly the dynamics of an
$n-$level subsystem. Another related objective is the study of {\it
heat current} characteristics in the spin-boson molecular junction
\cite{SegalR1,*SegalR2}. The single-bath spin-boson model displays a
rich dynamics with a complex phase diagram. Similarly, we expect
that the nonequilibrium version, with two harmonic baths of
different temperatures coupled to the TLS, will show complex
behavior for its heat current- temperature characteristics. Recent
results, obtained using an extension of the noninteracting blip
approximation to the nonequilibrium regime \cite{YelenaNIBA},
demonstrate rich behavior. Other problems that could be directed
with our method include plexcitonics systems, as the coupling
between surface plasmons and molecular excitons
 should be treated beyond the perturbative regime
\cite{Norlander}. Finally, we have discussed the calculation of
reduced density matrix and currents in the path-integral framework.
It is of interest to generalize these expressions and gain higher
order cumulants, for the study of current, noise, and fluctuation
relations in many-body out-of-equilibrium systems.


\begin{acknowledgments}
DS acknowledges support from an NSERC discovery grant. The work of
LS was supported by an Early Research Award of DS. Computations were
performed on the GPC supercomputer at the SciNet HPC Consortium
\cite{SciNet}. SciNet is funded by: the Canada Foundation for
Innovation under the auspices of Compute Canada; the Government of
Ontario; Ontario Research Fund - Research Excellence; and the
University of Toronto.
\end{acknowledgments}

\renewcommand{\theequation}{A\arabic{equation}}
\setcounter{equation}{0}  
\section*{Appendix A: Time-discrete Feynman-Vernon Influence functional}

With the discretization of the path, the influence functional takes
the form (\ref{eq:IBS}). The coefficients $\eta_{k,k'}$ were given
in \cite{QUAPI1,*QUAPI2} and we include them here for the
completeness of our presentation. The expressions are given here for
the case of a single boson bath with the initial temperature
$1/\beta_{ph}$ and the spectral function
$J_{ph}(\omega)=\pi\sum_{p}(\xi^{B}_{p})^2\delta(\omega-\omega_p)$,
$J_{ph}(\omega)=J_{ph}(-\omega)$,
\begin{widetext}
\bea \eta_{k,k'}&=&\frac{2}{\pi}\int_{-\infty}^{\infty} d\omega
\frac{J_{ph}(\omega)}{\omega^2}
\frac{\exp(\beta_{ph}\omega/2)}{\sinh(\beta_{ph}\omega/2)}
\sin^2(\omega\delta t/2)e^{-i\omega\delta t(k-k')} ,\,\,\,\,\,
0<k'<k<N
\nonumber\\
\eta_{k,k}&=&\frac{1}{2\pi}\int_{-\infty}^{\infty} d\omega
\frac{J_{ph}(\omega)}{\omega^2}
\frac{\exp(\beta_{ph}\omega/2)}{\sinh(\beta_{ph}\omega/2)}
\left(1-e^{-i\omega\delta t}\right),\,\,\,\, 0<k<N
\nonumber\\
\eta_{k,0}&=&\frac{2}{\pi}\int_{-\infty}^{\infty} d\omega
\frac{J_{ph}(\omega)}{\omega^2}
\frac{\exp(\beta_{ph}\omega/2)}{\sinh(\beta_{ph}\omega/2)}
\sin(\omega\delta t/4) \sin(\omega\delta t/2) e^{-i\omega(k\delta
t-\delta t/4)},\,\,\,\,\,\,\, 0<k<N
\nonumber\\
\eta_{N,k'}&=&\frac{2}{\pi}\int_{-\infty}^{\infty} d\omega
\frac{J_{ph}(\omega)}{\omega^2}
\frac{\exp(\beta_{ph}\omega/2)}{\sinh(\beta_{ph}\omega/2)}
\sin(\omega\delta t/4) \sin(\omega\delta t/2) e^{-i\omega(N\delta
t-k'\delta t -\delta t/4)},\,\,\,\,\,\,\, 0<k'<N
\nonumber\\
\eta_{N,0}&=&\frac{2}{\pi}\int_{-\infty}^{\infty} d\omega
\frac{J_{ph}(\omega)}{\omega^2}
\frac{\exp(\beta_{ph}\omega/2)}{\sinh(\beta_{ph}\omega/2)}
\sin^2(\omega\delta t/4)e^{-i\omega(N\delta t-\delta t/2)}
\nonumber\\
\eta_{0,0}&=&\eta_{N,N}=\frac{1}{2\pi}\int_{-\infty}^{\infty}
d\omega \frac{J_{ph}(\omega)}{\omega^2}
\frac{\exp(\beta_{ph}\omega/2)}{\sinh(\beta_{ph}\omega/2)}
\left(1-e^{-i\omega\delta t/2}\right) \eea
\end{widetext}



\bibliography{IF_Ref}
\end{document}